\documentclass[aps,prd,twocolumn,superscriptaddress,nofootinbib]{revtex4-2}
\usepackage{color}
\usepackage{graphicx}
\usepackage{booktabs}
\usepackage{appendix}
\usepackage{amsmath}
\usepackage{comment}
\usepackage{float}
\usepackage{inputenc}
\usepackage{fontenc}
\usepackage{multirow}
\usepackage[bookmarks,bookmarksnumbered]{hyperref}
\hypersetup{colorlinks = true,
            linkcolor = blue,
            anchorcolor =red,
            citecolor = blue,
            pdfauthor=author}

\usepackage[dvipsnames]{xcolor}
\usepackage{lineno}

\graphicspath{{plots/}}

\newcommand{\jetscape}{\textsc{jetscape}}
\newcommand{\martini}{\textsc{martini}}

\newcommand{\cujet}{\textsc{cujet}}
\newcommand{\matter}{\textsc{matter}}
\newcommand{\pythia}{\textsc{pythia}}
\newcommand{\fastjet}{\textsc{fastjet}}
\newcommand{\amy}{\textsc{amy}}
\newcommand{\dglv}{\textsc{dglv}}
\newcommand{\trento}{\textsc{trento}}
\newcommand{\vishnu}{\textsc{vishnu}}
\newcommand{\alphas}{\ensuremath{\alpha_\mathrm{s}}}
\newcommand{\raa}{\ensuremath{R_{AA}}}
\newcommand{\lbt}{\textsc{lbt}}
\newcommand{\adscft}{\textsc{ads-cft}}
\newcommand{\PbPb}{\ensuremath{\mathrm{Pb+Pb}}}
\newcommand{\pp}{\ensuremath{\mathrm{p+p}}}
\begin{document}

\title{
Comparing  the MARTINI and CUJET models for jet-quenching:\\ Medium modification of jets and jet substructure
}
    \author{Shuzhe Shi}
    \email{shuzhe.shi@stonybrook.edu}
    \affiliation{Center for Nuclear Theory, Department of Physics and Astronomy, Stony Brook University, Stony Brook, New York 11794–3800, USA}
    \affiliation{Department of Physics, McGill University, 3600 University street, Montreal, QC, Canada H3A 2T8}
    \author{Rouzbeh Modarresi Yazdi}
    \email{rouzbeh.modarresi-yazdi@mail.mcgill.ca}
    \affiliation{Department of Physics, McGill University, 3600 University street, Montreal, QC, Canada H3A 2T8}
    \author{Charles Gale}
    \affiliation{Department of Physics, McGill University, 3600 University street, Montreal, QC, Canada H3A 2T8}
    \author{Sangyong Jeon}
    \affiliation{Department of Physics, McGill University, 3600 University street, Montreal, QC, Canada H3A 2T8}

\begin{abstract}
Jets produced by the initial hard scattering in heavy ion collision events lose energy due to interactions with the color-deconfined medium formed around them: the quark-gluon plasma (QGP). Jet-medium interactions constitute an important theoretical and experimental field for studies of QGP, and various models with different assumptions have been proposed to describe them. A fair and direct comparison of these models require that all other aspects of the simulation be fixed, which is achieved in this work by relying on the \jetscape\, framework. We employ \jetscape\, to directly and comprehensively compare two successful energy loss models: \cujet\, and \martini. We compare the models with the results of measurements of jet spectra and substructure observables. With the strong coupling tuned separately, we find that the two models broadly agree with each other in nuclear modification factors for charged hadrons and jets with cone size $R=0.4$. Systematic differences are reported in fragmentation functions, jet shape, and cone size dependent jet \raa.
\end{abstract}

\maketitle
\section{Introduction}
Energetic partons moving through a strongly interacting plasma can undergo scattering with elements in 
the medium and lose energy. This phenomenon, known as ``jet-quenching'', has been observed 
and measured at major experimental facilities such as the BNL Relativistic Heavy Ion Collider (RHIC) and the CERN Large Hadron Collider (LHC), and is considered to be an 
important signal of the presence of the quark-gluon plasma (QGP) in the aftermath of a heavy ion collision. Beyond being a signal of 
the existence of the medium, however, jets are also considered ``hard probes'' of the QGP. 
This owes to the fact that jets are created at the moment of initial hard scattering 
and the medium is formed around them a short time ($\tau\lesssim 1 \,\mathrm{fm}/\mathrm{c}$) 
later. Therefore jets travel through the plasma as it evolves and are modified by it. 
In other words, the QGP carries its own probes. 

It is now accepted that gluon radiation resulting from jets interacting with medium particles is 
the dominant mechanism of jet energy loss. There has been an immense theoretical effort in 
modeling radiative energy loss of jets in a QGP, with different assumptions and approximations. 
For a recent review see~Ref.~\cite{Blaizot:2015lma}. In this work we focus on such two models of energy loss 
for jets at low virtuality: \martini~\cite{Schenke:2009gb}, which implements the \amy-McGill~\cite{Jeon:2003gi,Turbide:2005fk} formalism and \cujet~\cite{Xu:2014ica,Xu:2015bbz,Shi:2018izg,Shi:2018lsf} which employs the \dglv~\cite{Gyulassy:1999zd,Gyulassy:2000er,Djordjevic:2003zk} radiative rates. 
 
A previous comparison of various perturbative QCD (pQCD) based radiative energy loss formalisms, including \amy\, 
and \dglv, was performed in a static QGP ``brick''~\cite{Armesto:2011ht}. The focus was placed on the radiative rates and their specific assumptions both in the physics at the stage of derivation 
and on the details of implementation.
Later on, the JET Collaboration~\cite{JET:2013cls} compiled the result of different models, 
and obtained global-fitted value for the scaled jet transport parameter $\hat{q}/T^3$.
Comparison of the energy loss models in realistic simulations, however, provides its own challenges related to the different modeling choices: the initial condition of the 
hydro evolution, the temperature parametrization(s) of viscosities, the initial jet distribution and so on. 
Thus a careful comparison would need to minimize, or at least control, the possible consequences of these choices. 

In this work, we use the \jetscape\, framework~\cite{Putschke:2019yrg}, developed specifically 
to address these difficulties. The modular approach of \jetscape\, allows for fixing all 
aspects of the simulation except for those of  the specific model under study. In this way, the 
energy loss formalisms of interest see the same initial conditions, jet distribution, hydrodynamic 
history, and hadronization mechanisms.
For jet energy loss and analysis, 
\jetscape\, is shipped with \matter~\cite{Majumder:2013re,Cao:2017qpx} used for vacuum and 
in-medium high virtuality final state showers and \pythia~\cite{Bierlich:2022pfr} for the hard 
scattering generation, initial state shower, and final fragmentation to hadrons. The included low 
virtuality energy loss modules are \martini, \lbt~\cite{He:2015pra,Cao:2016gvr}, and \adscft~\cite{Casalderrey-Solana:2014bpa}. The \jetscape\, 
Collaboration has previously presented comparative studies of jet energy loss for those models 
which serve to illustrate the flexibility and power of the framework~\cite{JETSCAPE:2017eso,Park:2019sdn,JETSCAPE:2018vyw}. For a full description of the default packages and models implemented in \jetscape\, we refer the interested reader to Ref$.$~\cite{Putschke:2019yrg}. 

Importantly, \cujet\, is not a standard \jetscape\, package: in this work we incorporated it 
into the \jetscape\, event flow as an available low virtuality energy loss module. 
This required recasting the deterministic, standalone implementation of \cujet\, 
into a Monte Carlo version. The incorporation of \cujet\, into \jetscape\, then allows, 
for the first time, to have a direct comparison of the \cujet\, and \martini\, energy 
loss formalisms with great control over all other aspects of evolution\footnote{From here on, ``\cujet'' will refer to the Monte Carlo implementation in \jetscape\, and ``standalone \cujet'' will refer to the deterministic, standalone \cujet\, package.}.

In this paper, part one of a two-paper series, we focus on jet spectra and sub-structure 
in order to compare \cujet\, and \martini. The study of electromagnetic probes 
(specifically jet-medium photons) resulting from the two energy loss formalisms is left 
to the second installment of this work. The outline of the paper is as follows: 
we discuss the physics of energy loss in the respective implementation of \cujet\, and \martini\, in 
Sec$.$~\ref{sec.eloss}. Section.~\ref{sec.comparison.static} provides our comparisons of 
the two modules in a static QGP brick while Sec.~\ref{sec.comparison.hydro} presents the 
results of embedding the two models in a realistic viscous hydrodynamic simulation. We present 
our conclusions as well as an outlook of future work in Sec$.$~\ref{sec.conclusion.outlook}.

\section{Energy Loss}\label{sec.eloss}
In this section, we describe details of two energy loss models under study, focusing especially on the differences and show results in a QGP brick of fixed temperature. We then discuss, qualitatively, what difference in jet substructure would be expected. The starting point is the collisional energy loss channel, where the two models are 
the most similar. A discussion on radiative energy loss then follows.

\subsection{Collisional energy loss}
In both the \cujet\ and \martini\ models, the elastic scattering process is implemented as the 
leading order $2\rightarrow 2$ scattering channels between gluons, quarks and antiquarks. 
Both models compute the elastic rates in the $t$-channel dominance approximation where it is assumed that the Mandelstam-t channel is the main contribution to the 
total scattering cross section. \cujet\, then takes the differential cross section to be 
\begin{align}
    \frac{\mathrm{d} \sigma_{i,j}}{\mathrm{d} t} = 
    \frac{2\pi \alpha_s^2}{(t+m^2_D)t} c_{i,j}\,,
    \label{eq.elastic_diff}
\end{align}
where the color factors are $c_{i,j} = 4/9$, $1$, and $9/4$ for $\{i,j\} = \{q,q\}$, $\{q,g\}$, 
and $\{g,g\}$ and similarly for anti-quarks. Finally, $m_D$ is the Debye screening mass given by
\begin{equation}
   m_D^2 = g_s^2 T^2 (2N_c+N_f)/6
   \label{eq.Debye.Screening.Mass}
\end{equation}
with $T$ being the local temperature and $N_c$ and $N_f$ denoting, respectively, the number of colors ($N_c=3$) and flavors ($N_f=3$) under study. 

In \cujet, the total rate of elastic scattering is given by
\begin{align}
\Gamma^{\cujet}_\mathrm{ela}(p,T) = \sum_{j} d_j \int \frac{\mathrm{d}^3\mathbf{k}}{(2\pi)^3}\,
	f_{j} (T,\mathbf{k}) \int \mathrm{d}{t} \frac{\mathrm{d}\sigma_{i,j}}{\mathrm{d}{t}} \,,
    \label{eq.cujet.elastic}
\end{align}
with $d_j$ being the degeneracy and $f_{j}$ the distribution function of particle $j$.

In \martini, the scattering rates take into account the Pauli blocking and Bose stimulation effects for the recoil parton~\cite{Qin:2007rn,Qin:2008ea},
\begin{multline}
    \frac{\mathrm{d}\Gamma^{\martini}_\mathrm{ela}}{\mathrm{d}\omega}(E,\omega,T) = \frac{d_k}{(2\pi)^3} \frac{1}{16 E^2} \int_0^p \mathrm{d}q\times\,\nonumber\\
    \int_{\frac{q-\omega}{2}}^\infty \mathrm{d}k\,\theta(q-|\omega|)
    \int_0^{2\pi}\frac{\mathrm{d}\phi_{kq|pq}}{2\pi} |\mathcal{M}|^2 f(k,T)\left[1\pm f(k',T)\right]
\end{multline}
where $d_k$ is the degeneracy factor of the thermal parton, $q$ the exchanged momentum, $p$ the momentum of the 
incoming jet and $k$ the momentum of the medium particle. The angle $\phi_{kq|pq}$ measures the angle between 
the $\mathbf{k}\times\mathbf{q}$ and $\mathbf{p}\times\mathbf{q}$ planes~\cite{Schenke:2009ik} and $\mathcal{M}$ is the matrix element of the process where the hard-thermal-loop (HTL) gluon propagator is used to cure the infrared divergences~\cite{Schenke:2009gb}.

Other than the above, \martini\, also includes ``conversion'' channels where via soft fermion exchange, the incoming jet, $q(\bar{q})$ or $g$, is converted to a $g$ or $q(\bar{q})$ respectively. These processes are also dominated by their respective Mandelstam-t channel diagrams, and their rates are given by~\cite{Schenke:2009gb}
\begin{align}
    \frac{\mathrm{d}\Gamma^{\mathrm{conv}}_{q\to g}}{\mathrm{d}p} &= C_\mathrm{F} \frac{2\pi\alpha^2_s T^2}{3p}\left(\frac{1}{2}\ln{\frac{pT}{m^2_q}}-0.36149\right)\nonumber\\
    \frac{\mathrm{d}\Gamma^{\mathrm{conv}}_{g\to q}}{\mathrm{d}p} &= N_f \frac{N_c}{N^2_c-1}\frac{\mathrm{d}\Gamma^{\mathrm{conv}}_{q\to g}}{\mathrm{d}p}
    \label{eq:parton_conv_processes}
\end{align}
where $p$ is the momentum of the incoming jet. The momentum of the outgoing parton is also $p$ in a conversion process. In addition, $N_c=3$ is the number of colors, $C_\mathrm{F} = \frac{N_c^2-1}{2N_c} = 4/3$ the Casimir factor for quarks, and $N_f=3$ is the number of flavors. Finally, $m_q$ is the thermal mass of the quark and it is given by
\begin{equation}
    m^2_{q} = g^2_s T^2 /6.
    \label{eq.thermal.quark.mass}
\end{equation}
In \martini\, the strong coupling of the conversion channels is set equal to that of other elastic processes.

\subsection{Radiative energy loss}

    \begin{figure*}[!tp]\centering
        \includegraphics[width=0.9\textwidth]{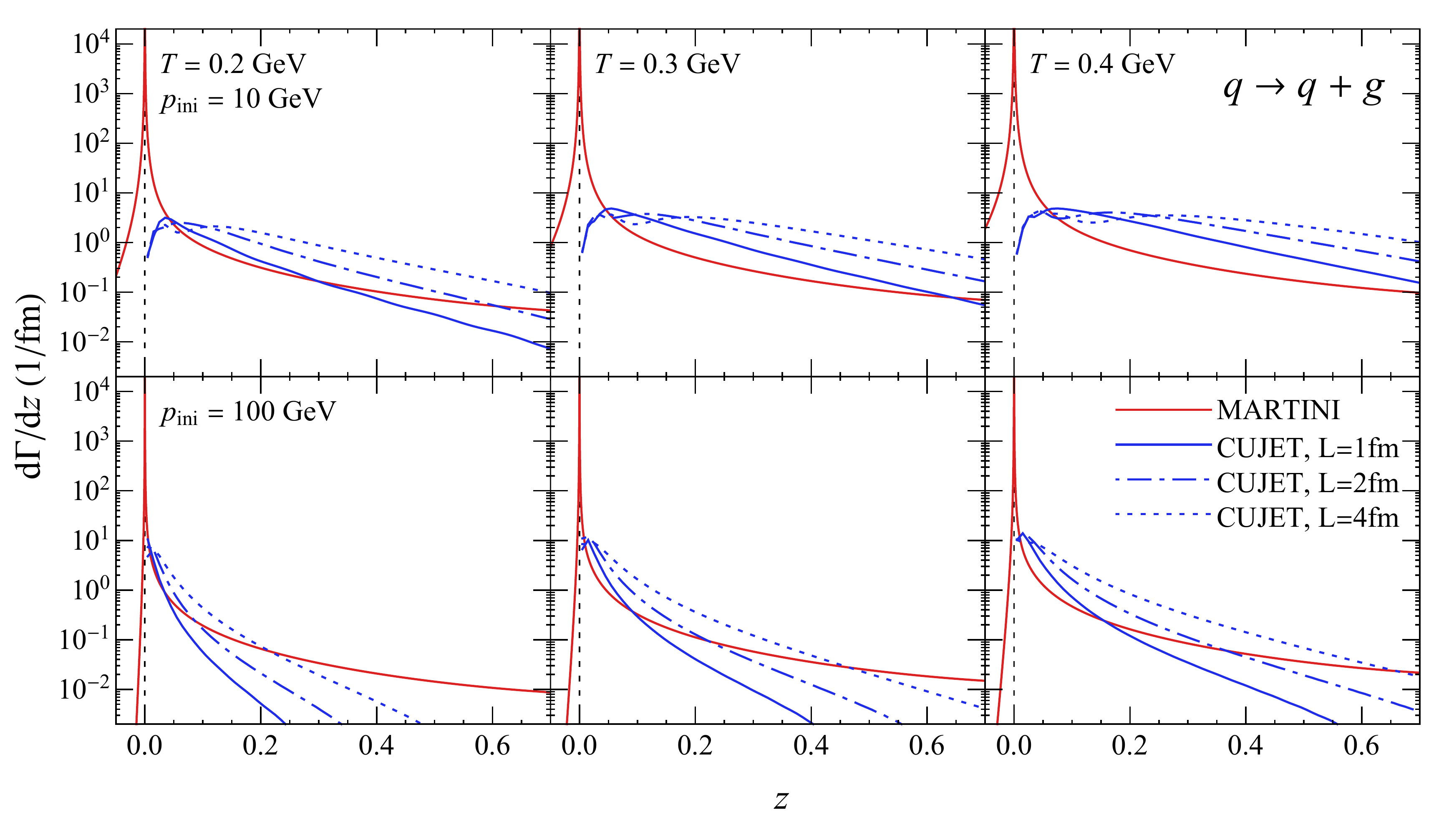}
        \caption{Quark splitting rates $\mathrm{d}\Gamma_{q\to qg}/\mathrm{d}z$ as functions of energy loss ratio, $z\equiv p_g / p_\mathrm{ini}$.
        From left to right panels correspond to the temperature $T=0.2$, $0.3$, and $0.4$~GeV, respectively, whereas top(bottom) panels are for quarks with initial momentum $p_\text{ini}=10$ $(100)$ GeV. The red curves are splitting rates used in \martini~[Eq.~\ref{eq.splitting_rate_martini}], which are independent of path length, and the blue curves are for those of \cujet~[Eq.~\ref{eq.splitting_rate_cujet}] at path length $L = 1$ (solid), $2$ (dash-dotted), and $4$ (dotted) fm.
        \label{fig.rate}}
    \end{figure*}
    
Despite  their similarity in treating collisional energy loss, the \martini\ and \cujet\ 
models are fundamentally different when describing inelastic collisions. \martini\ 
adopts the \amy\ formalism which is evaluated to all orders of opacity and assumes an 
infinite thermal medium, while  \cujet\, computes the rates up to  first order in 
the opacity expansion in the \dglv\ formalism but accounts for the finite medium size. 
 
In this section, we will provide details for the implementation of radiative processes in 
both models and then compare their properties.

\subsubsection{Parton splitting in \amy-\martini}
The \martini\, simulation framework~\cite{Schenke:2009gb} takes the \amy\,  formalism~\cite{Arnold:2001ba,Arnold:2001ms,Arnold:2002ja}, in which the inelastic splitting rates  $\mathrm{d} \Gamma_{i\to jk}/\mathrm{d}z$ are formulated as functions of in-coming momentum $p$ and energy-loss ratio $z \equiv p_\text{out}/p$ 
    \begin{align}
        \begin{split}
        \frac{\mathrm{d} \Gamma^{\amy}_{i\to jk}}{\mathrm{d} z} (p,z) =\;& 
        	\frac{\alpha_s P_{i \to jk}(z)}{[2p\,z(1{-}z)]^2} \bar{f}_j(z\,p)\, \bar{f}_k((1-z)p) 
        \\&\times	
        	\int \! \frac{\mathrm{d}^2 \mathbf{h}_{\perp}}{(2\pi)^2} ~\text{Re} \left[ 2\mathbf{h}_{\perp} \cdot \mathbf{g}_{(z,p)}(\mathbf{h}_{\perp}) \right]\;,
        \end{split}\label{eq.splitting_rate_martini}
    \end{align}
where $\bar{f}=1+f$($\bar{f}=1-f$), with $f$ being the Bose--Einstein~(Fermi--Dirac) distribution for outgoing gluons~(quarks), which accounts for the Bose enhancement~(Pauli blocking) effect and reflect how \amy\, models the thermal medium: as a weakly coupled collection of gluons and quarks. The $P_{i\to jk}(z)$ are the Dokshitzer--Gribov--Lipatov--Altarelli--Parisi~(DGLAP) \cite{Gribov:1972ri,Lipatov:1974qm,Altarelli:1977zs,Dokshitzer:1977sg} splitting functions
\begin{align}
\begin{split}
    P_{g \to gg}(z) =\;& 2 C_\mathrm{A} \frac{[1-z(1{-}z)]^2}{z(1{-}z)}\;,\\
    P_{q \to qg}(z)  =\;& C_\mathrm{F} \frac{1+(1{-}z)^2}{z}\;, \\
    P_{g \to q\bar{q}}(z) =\;& \frac{1}{2} \left(z^2+(1{-}z)^2\right) \;,
\end{split}
\end{align}
with the Casimir factor $C_\mathrm{A} = N_c = 3$.

The function $\mathbf{g}_{(z,p)}(\mathbf{h}_{\perp})$, which encodes the current-current correlator, satisfies the following integral equation:
\begin{align}
\label{eq:AMY}
\begin{split}
    2\mathbf{h}_{\perp} =\;& i \delta E(z,p,\mathbf{h}_{\perp}) \mathbf{g}_{(z,p)}(\mathbf{h}_{\perp}) 
    + \int \frac{\mathrm{d}^2\mathbf{q}_{\perp}}{(2\pi)^2}~\bar{C}(q_\perp)   \\
    &\times \Big\{ C_{1} [ \mathbf{g}_{(z,p)}(\mathbf{h}_{\perp}) - \mathbf{g}_{(z,p)}(\mathbf{h}_{\perp} -\mathbf{q}_{\perp}) ]  \\
    & + \, C_{z} [\mathbf{g}_{(z,p)}(\mathbf{h}_{\perp}) - \mathbf{g}_{(z,p)}(\mathbf{h}_{\perp} -z\mathbf{q}_{\perp}) ]   \\
    & + \, C_{1-z} [\mathbf{g}_{(z,p)}(\mathbf{h}_{\perp}) - \mathbf{g}_{(z,p)}(\mathbf{h}_{\perp} -(1{-}z)\mathbf{q}_{\perp}) ] \Big\} \, 
\end{split}
\end{align}
where $\mathbf{h}_{\perp}$ determines the of collinearity of the outgoing particles ($\mathbf{h}_{\perp} = (\mathbf{p}_{\mathrm{out}} \times \hat{\mathbf{p}})\times \hat{\mathbf{p}}_{\parallel}$) and the energy difference between the initial and final states, $\delta E(z,p,\mathbf{h}_{\perp})$, is given by
\begin{eqnarray}
\delta E(z,p,\mathbf{h}_{\perp}) = \frac{\mathbf{h}_{\perp}^2}{2p\,z(1{-}z)} +M_{\rm eff}(z,p)\;,
\end{eqnarray}
where $M_{\rm eff}(z,p)$ is given in terms of the asymptotic masses $m^2_{\infty,(1,z,1-z)}$ of the particles with momentum fractions $1,z,1-z$ as
\begin{align}
M_{\rm eff}(z,p)=  \frac{m^2_{\infty,(z)}}{2zp}  + \frac{m^2_{\infty,(1{-}z)}}{2(1{-}z)p} -\frac{m^2_{\infty,(1)}}{2p}.
\end{align}
For the asymptotic masses we use the leading order results given by
\begin{align}
    m^2_{\infty,g}=& \frac{m_{D}^2}{2}= \frac{g_s^2T^2}{6}\left( C_\mathrm{A} +\frac{N_\mathrm{f}}{2}\right),\nonumber\\
    m^2_{\infty,q}=& 2 m^2_q = C_\mathrm{F} \frac{g_s^2 T^2}{4}\,,
\end{align}
with the Debye screening mass and the thermal quark mass given by Eq.~\ref{eq.Debye.Screening.Mass} and Eq.~\ref{eq.thermal.quark.mass}, respectively. The color factors are given by
\begin{align}\begin{split}
C_{1}=\;&\frac{1}{2} \Big( C^{R}_{z} + C^{R}_{1-z} - C^{R}_{1} \Big)\;, \\
C_{z}=\;&\frac{1}{2} \Big( C^{R}_{1-z} + C^{R}_{1} - C^{R}_{z} \Big)\;,  \\
C_{1-z}=\;&\frac{1}{2} \Big( C^{R}_{1} + C^{R}_{z} - C^{R}_{1-z} \Big)\;, 
\end{split}\end{align}
where $C^{R}_{(1,z,1-z)}$ denote the Casimir of the representation of the particle carrying momentum 
fraction $1,z,1-z$, i.e. $C^{R}=C_\mathrm{F}$ for quarks and $C^{R}=C_\mathrm{A}$ for gluons. Since 
the color factors have been factored out, the rate $\bar{C}(q)$ in Eq.~(\ref{eq:AMY}) denotes the 
elastic scattering rate stripped of its color factor. 

\begin{figure*}[!tp]\centering
    \includegraphics[width=0.9\textwidth]{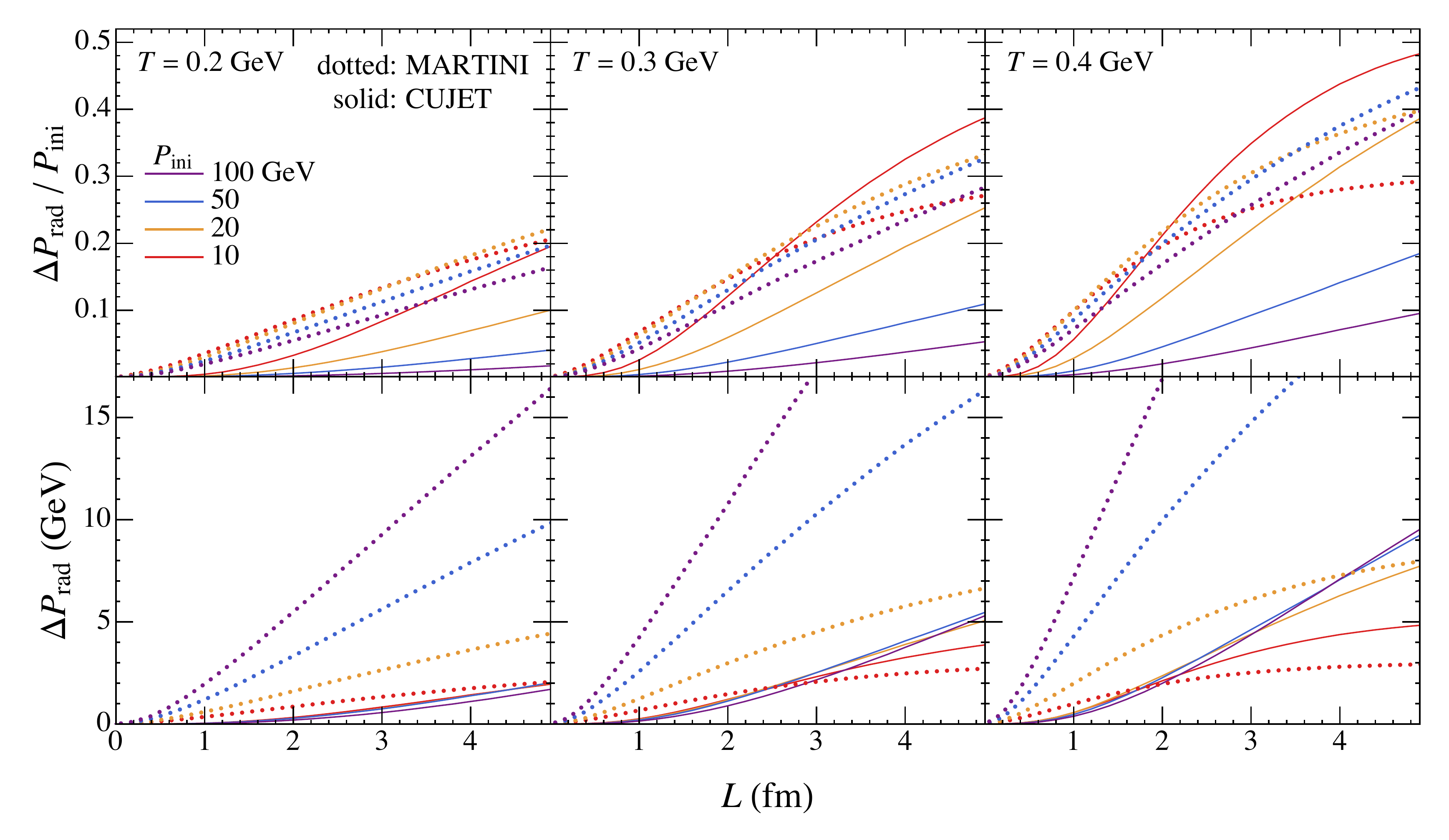}
    \caption{Quark radiation momentum loss $\Delta p_\text{rad}$ (bottom) and momentum-loss-ratio $\Delta p_\text{rad}/ P_\text{ini}$ (top) as functions of path length $L$. From left to right panels correspond to the temperatures $T=0.2$, $0.3$, and $0.4$~GeV, respectively, whereas red, orange, blue, and purple curves are for quarks with initial momenta $p_\text{ini}=10$, $20$, $50$, $100$ GeV.  The solid(dotted) curves are for \cujet(\martini).
\label{fig.radel}}
\end{figure*}
\begin{figure*}[!hbt]\centering
        \includegraphics[width=0.8\textwidth]{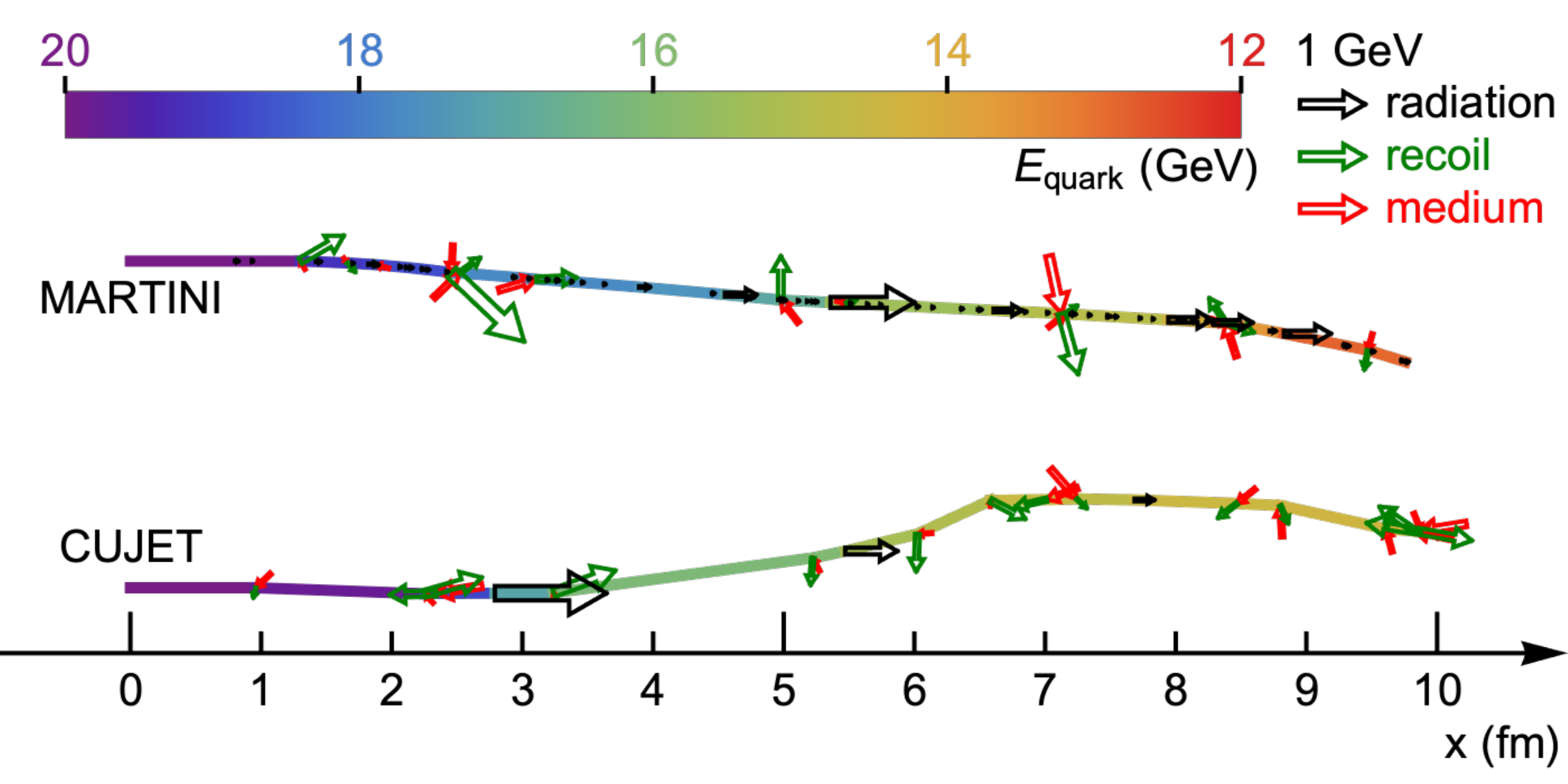}
    \caption{History of two hard partons with initial energy $E_\mathrm{ini}=20$~GeV and final energy $E_\mathrm{ini}\approx 12$~GeV, evolved according to the \martini\,(upper) and \cujet\,(lower) energy loss models, respectively. The horizontal (vertical) coordinate indicates the hard parton position along(perpendicular to) the initial direction. The color of the ``tube'' stands for the energy of the energetic quark. Black, green, and red arrows respectively stand for the momentum vectors of the radiated gluons, recoil partons, and incoming medium partons.\label{fig.configuration}}
\end{figure*}

Finally, \martini\, uses the leading order perturbative expression for the running of \alphas\, and takes the scale to be the average momentum transfer squared $\langle p^2_\perp \rangle$
\begin{align}
    \alpha_s = \alpha_s(\langle p_\perp^2\rangle) = \frac{4\pi}{9\ln(\langle p_\perp^2\rangle/\Lambda_\mathrm{QCD}^2)}
    \label{eq.martini.running_coupling}
\end{align}
with $\Lambda_{\mathrm{QCD}}$ set to $200$ MeV. The strong coupling for the radiative and elastic processes are allowed to run with different scales and here we take them to be proportional to the average momentum transfer of the process
\begin{align}
    \langle p_\perp^2\rangle = \left\{
\begin{array}{ll}
\kappa_r^2 \,\sqrt{\hat{q}\, p}& \text{radiative process,} \\
\kappa_e^2 \,\hat{q}\, \lambda_\mathrm{mfp}     & \text{elastic process,}
\end{array}
    \right.
    \label{eq.running_scales_martini}
\end{align}
where $\kappa_r$ and $\kappa_e$ are the constants of proportionality to be determined via fits to experimental data. For the radiative energy loss channel, the scale is chosen to be the quartic root of the product of the average momentum transfer per unit length ($\hat{q}$) and the incoming parton momentum $p$~\cite{Young:2012dv}. The elastic energy loss channel's renormalization scale is taken to be proportional to the product of the mean free path of the incoming parton ($\lambda_{\mathrm{mfp}}$) and $\hat{q}$~\cite{Park:2021yck}. To use $\hat{q}$ and $\lambda_{\mathrm{mfp}}$ we compute them using the elastic scattering rate, which can be derived analytically~\cite{JET:2013cls}:
\begin{align}
    \hat{q} &= \int^{q_{\mathrm{max}}} d^2\mathbf{q}_{\perp}\,\mathbf{q}^2_{\perp}\,\frac{d\Gamma_{\mathrm{elas.}}}{d^2\mathbf{q}_{\perp}}\,, \nonumber\\
    \lambda_{\mathrm{mfp}} &= \Gamma_{\mathrm{elas.}}^{-1}\,, \nonumber\\
    \Gamma_{\mathrm{elas.}} &= \int^{q_{\mathrm{max}}}_{q_{\mathrm{min}}}\,d^2\mathbf{q}_{\perp}\,\frac{d\Gamma_{\mathrm{elas.}}}{d^2\mathbf{q}_{\perp}}\,,
\end{align}
where the elastic collision rate is given by
\begin{equation}
    \frac{d\Gamma_{\mathrm{elas.}}}{d^2\mathbf{q}_{\perp}} = \frac{C_{R}}{(2\pi)^2} \frac{g^2 m^2_D T}{\mathbf{q}^2_{\perp}(\mathbf{q}^2_{\perp}+m^2_D)}.
\end{equation}
Thus the final expressions for the mean free path and the average momentum transfer per unit length are~\cite{Park:2021yck}
\begin{align}
\hat{q} =\;& 
    C^R \alpha_{s,0} m^2_D T \ln(1+q^2_\mathrm{max}/m_D^2)\,,\\
\lambda_\mathrm{mfp} =\;&
    \bigg(C^R \alpha_{s,0} T \ln\frac{1+m_D^2/q^2_\mathrm{max}}{1+m_D^2/q^2_\mathrm{min}} \bigg)^{-1}\,.
\end{align}
Here, $q^2_\mathrm{max} = 2\, p\, k_\mathrm{th} = 6\,p\,T$ is the maximum momentum transferred, where $k_\mathrm{th} = 3T$ is the average momentum of the in-medium soft particles. The infrared cut-off is set to $q_\mathrm{min} = 0.05\,T$ to be consistent with the minimum momentum transfer used for the calculations of the total elastic
rates~\cite{Schenke:2009ik} implemented in \martini. Furthermore, the maximum running of \alphas\, for either radiative or elastic collisions in \martini\, has an upper bound of $0.42$ while the minimum allowed value for elastic scattering is set to $0.15$.

\subsubsection{Parton splitting in \dglv-\cujet}
The \cujet\, energy loss model~\cite{Buzzatti:2011vt,Xu:2014ica,Xu:2014tda,Xu:2015bbz,Shi:2018izg,Shi:2018lsf} takes into account the finite-size medium and employs the \dglv\, formalism~\cite{Gyulassy:1999zd,Gyulassy:2000er,Djordjevic:2003zk} to compute the inelastic parton splitting. The latter explicitly depends on not only the in-coming energy and energy loss ratio, but also on the time since the last splitting ($\tau$),
\begin{align}
\begin{split}
&\frac{\mathrm{d}\Gamma^{\dglv}_{i \to g i}}{\mathrm{d}z}(p,z,\tau) \\
=\;& 
 	\frac{18 C^R_{i}}{\pi^2} \frac{4+N_f}{16+9N_f} \rho(T)
\\&	\times
	\int{\mathrm{d}^2\mathbf{k}_{\perp}} \Bigg\{
	\frac{1}{z_+} \left| \frac{\mathrm{d}z_+}{\mathrm{d}z} \right| 
	\alpha_s \Big( \frac{\mathbf{k}_{\perp}^2}{z_+ - z_+^2} \Big)
\\ & \times
	\int \frac{ \mathrm{d}^2\mathbf{q}_{\perp}}{\mathbf{q}_\perp^2} \Bigg[
	\frac{ \alpha_s^2(\mathbf{q}_{\perp}^2)}{\mathbf{q}_\perp^2 + m_D^2} 
	\frac{-2}{(\mathbf{k}_{\perp}-\mathbf{q}_{\perp})^2+\chi^2}
\\ & \times
	 \bigg( \frac{\mathbf{k}_{\perp}\cdot(\mathbf{k}_{\perp}-\mathbf{q}_{\perp})}{\mathbf{k}_{\perp}^2+\chi^2} - \frac{(\mathbf{k}_{\perp}-\mathbf{q}_{\perp})^2}{(\mathbf{k}_{\perp}-\mathbf{q}_{\perp})^2+\chi^2} \bigg)
\\& \times
	\bigg(1-\cos\bigg(\frac{(\mathbf{k}_{\perp}-\mathbf{q}_{\perp})^2+\chi^2}{2 z_+ p} \tau\bigg)\bigg) \Bigg] \Bigg\}\;\;,
\end{split}
\label{eq.splitting_rate_cujet}
\end{align} 
where $i=q$ or $g$, the gluon plasmon mass $ m_g(T) = m_D(T) / \sqrt{2}$, while $\chi^2(T) = M^2 z_+^2+m_g^2(1-z_+)$ regulates the soft collinear divergences in the color antennae and controls 
the LPM phase. Note that $M$ is the mass of the quark which in this work, given our focus on gluon and light quarks, is set to zero.
\cujet\, models the medium as a well-separated assembly of Debye-screened scattering centers and estimates the soft parton number density $\rho(T)$ from entropy density 
$\rho=s/4$, where the relation between entropy density and temperature is given by the 
\texttt{s95p-PCE} equation of state. The gluon fractional energy $z$ and fractional plus-momentum 
$z_+$ are connected by $z_+ = z[1+\sqrt{1-(k_\perp/z p)^2}]/2$. The limit of integration is 
$|\mathbf{q}_\perp| \leq q_\mathrm{max} = \sqrt{6\,p\,T}$, 
$|\mathbf{k}_\perp| \leq z p$. $\mu$ is the gluon thermal mass 
which satisfies the self-consistent equation:
\begin{align}
m_D^2(T) = 4\pi\, \alpha_s(m_D^2)\, T^2\, (1+N_f/6) \,.
\label{eq.cujet.gluon_thermal_mass}
\end{align}
In \cujet, the running coupling utilizes the one-loop result with a soft plateau:
\begin{equation}
    \alpha_s(Q^2) = \begin{cases}
        \frac{4\pi}{9\ln(Q^2/\Lambda_\mathrm{QCD}^2)} \,, &
	            Q > \Lambda_\mathrm{QCD} \,e^{\frac{2\pi}{9\alpha_{\max}}} \,,\\~\\
        \alpha_{\max} \,, & 
	            Q \leq \Lambda_\mathrm{QCD} \,e^{\frac{2\pi}{9\alpha_{\max}}} \,.
\end{cases}
\label{eq.cujet.running_coupling}
\end{equation}
where $\alpha_{\max}$ is a parameter of the model to extracted from a fit to data~(see Appendix.~\ref{appendix_A} for more detail). 
It should be noted that the $g\to q\bar{q}$ splitting channel has been neglected in \cujet,  
 but is not expected to cause any phenomenologically measurable effects 
in heavy-ion collisions.

\section{Comparison in a static medium}\label{sec.comparison.static}
We start with a direct comparison of the \martini\ and \cujet\ splitting rates. This is shown in 
Fig.~\ref{fig.rate} where one can immediately observe remarkable differences between the 
two radiative rates of the two models. First, the \martini\, rates contain both emission ($z>0$) and 
absorption($z<0$) sectors and they peak at $z=0$. \cujet\ rates, on the other hand, are restricted to $z>0$ and peak at finite $z$. Second, the \cujet\ and \martini\ rates exhibit different 
dependence on the initial jet energy: while at the large-$z$ limit, \cujet\ rates always
decay more rapidly than the \martini\ ones, they also exhibit a strong momentum dependence (the large-$z$ tail for a quark with $p_\mathrm{ini}=100$~GeV decays faster than that of a $p_\mathrm{ini}=10$~GeV quark). This is in contrast to \martini\, rates which remain mostly flat in the large $z$ limit. As a consequence, these two models predict different energy sensitivities and different parton distributions as a result of the in-medium quenching.
To further investigate the characteristics of \cujet\ and \martini\ energy loss mechanisms, we perform a brick test of the two models in Fig$.$~\ref{fig.radel} by injecting a high-momentum ($p_\mathrm{ini}$) quark into a static, homogeneous QGP brick at constant temperature ($T$) and finite length ($L$). The parton is then evolved according to the \cujet\ and \martini\ models respectively and final momentum ($p_\mathrm{fin}$) after the evolution is measured. After averaging over the Monte Carlo events, we obtain the expectation of radiation energy loss $\Delta p_\mathrm{rad} \equiv p_\mathrm{ini} - \langle p_\mathrm{fin}\rangle$ as a function of path length for various initial energies and brick temperatures. For this comparison, we only allow radiative energy loss and turn off the collisional channels. As expected, we observe a clear difference between these two models. Especially, the net momentum-loss ($\Delta p_\mathrm{rad}$) is insensitive to $p_\mathrm{ini}$ in \cujet\ whereas the \martini\ calculation predicts a momentum-loss-ratio ($\Delta p_\mathrm{rad}/p_\mathrm{ini}$) that is insensitive to the initial momentum. Furthermore, the temperature and path length dependencies are also different. After a quantitative comparison, we find the empirical relations 
\begin{align}
    \Delta p_\text{rad}^{[\cujet]} \propto p^0 T^3 L^2\nonumber\,,\\
    \Delta p_\text{rad}^{[\martini]} \propto p^1 T^1 L^1\,.
    \label{eq.energy.loss.empirical}
\end{align}

A more intuitive comparison is shown in Fig.~\ref{fig.configuration}, where we randomly select 
two events --- evolved according to \martini\ and \cujet\ mechanisms, respectively --- with the same initial energy~($E_\mathrm{ini}=20$~GeV) and similar final energy~($E_\mathrm{fin}\approx 12$~GeV) of the leading parton
and plot the history of elastic and inelastic scatterings occurred inside a QGP brick with temperature
$T=0.3$~GeV and thickness $L=10$~fm. Distributions of the radiated gluons are visibly different: the
quark, when evolved by \martini\ emits and absorbs many more soft gluons as opposed to when \cujet\ governs the evolution. 
We also observe that \martini\  predicts more radiative energy loss compared to \cujet, which agrees with the comparison in Fig$.$~\ref{fig.radel}. A note of caution in interpreting Fig.~\ref{fig.configuration} is in order: given \martini's propensity to lose more energy via radiation, our event selection requirement of similar final state energy is translated into selection of events with more elastic energy loss in \cujet\, in order to compensate for the total energy loss. In other words, the plot does not necessarily mean that \cujet\, \textit{generally} predicts more elastic energy loss than \martini.

To conclude, the \cujet\ and \martini\ energy loss models predicts different characteristics of the in-medium jet-related parton distributions, which may lead to observable effects in the substructure of jets created in heavy-ion collisions. In the next section we investigate the phenomenological differences
of the two models in the context of a realistic simulation of heavy ion collisions.

\begin{figure*}[!ht]\centering
    \includegraphics[width=0.9\textwidth]{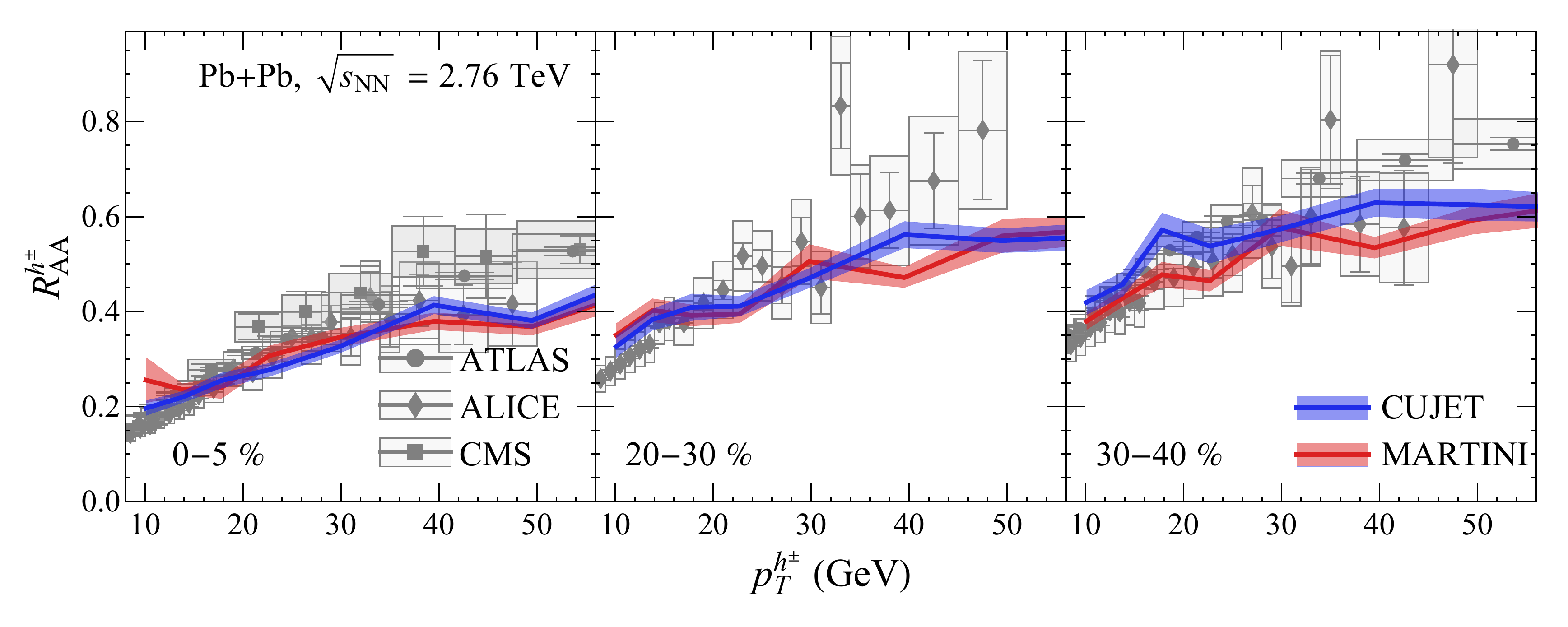}
    \caption{Charged hadron nuclear modification factor ($R_{AA}^{h^\pm}$) versus transverse momentum ($p_T^{h^\pm}$) in $\mathrm{Pb}+\mathrm{Pb}$ collisions at beam energy $\sqrt{s_{NN}}=2.76$~TeV. From left to right corresponds to $0-5\%$, $20-30\%$, and $30-40\%$ centrality range. Red and blue curves represent simulation results using \matter+\martini\, and \matter+\cujet, respectively. Both theoretical calculation and experimental measurements~\cite{ALICE:2012aqc,ATLAS:2015qmb,CMS:2012aa} are for charged hadrons with pseudo-rapidity $|\eta|<1$.
    \label{fig.AA.chRAA}}
    \includegraphics[width=0.9\textwidth]{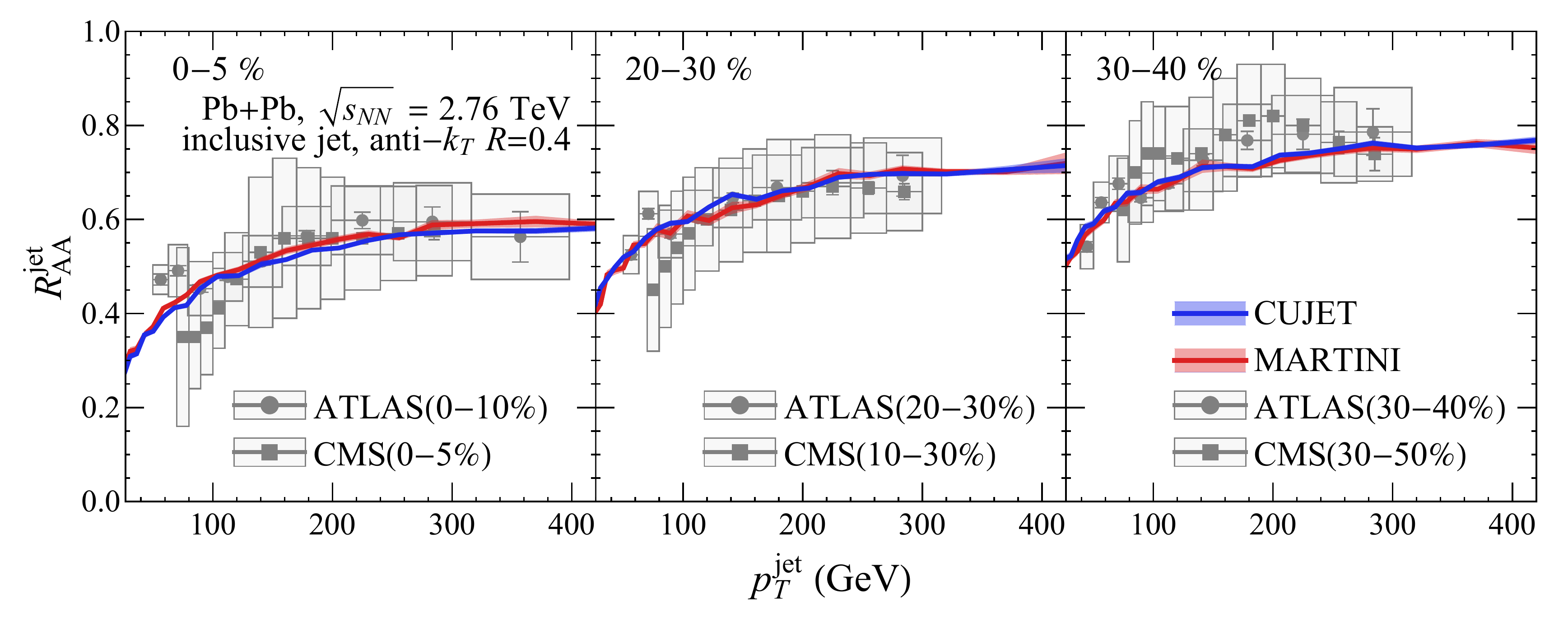}
    \caption{Same as Fig.~\protect{\ref{fig.AA.chRAA}} but for inclusive jet $R_{AA}$.
        Simulation include jets within pseudo-rapidity range $|\eta_\mathrm{jet}| < 2$, whereas
        experimental results are respectively for $|\eta_\mathrm{jet}| < 2.1$ in ATLAS~\cite{ATLAS:2014ipv} and $|\eta_\mathrm{jet}| < 2$ in CMS~\cite{CMS:2016uxf} measurements.
\label{fig.AA.jetRAA}}
\end{figure*}

\section{Comparison in Realistic Simulations}
\label{sec.comparison.hydro}

\begin{figure*}[!hbtp]\centering
    \includegraphics[width=0.9\textwidth]{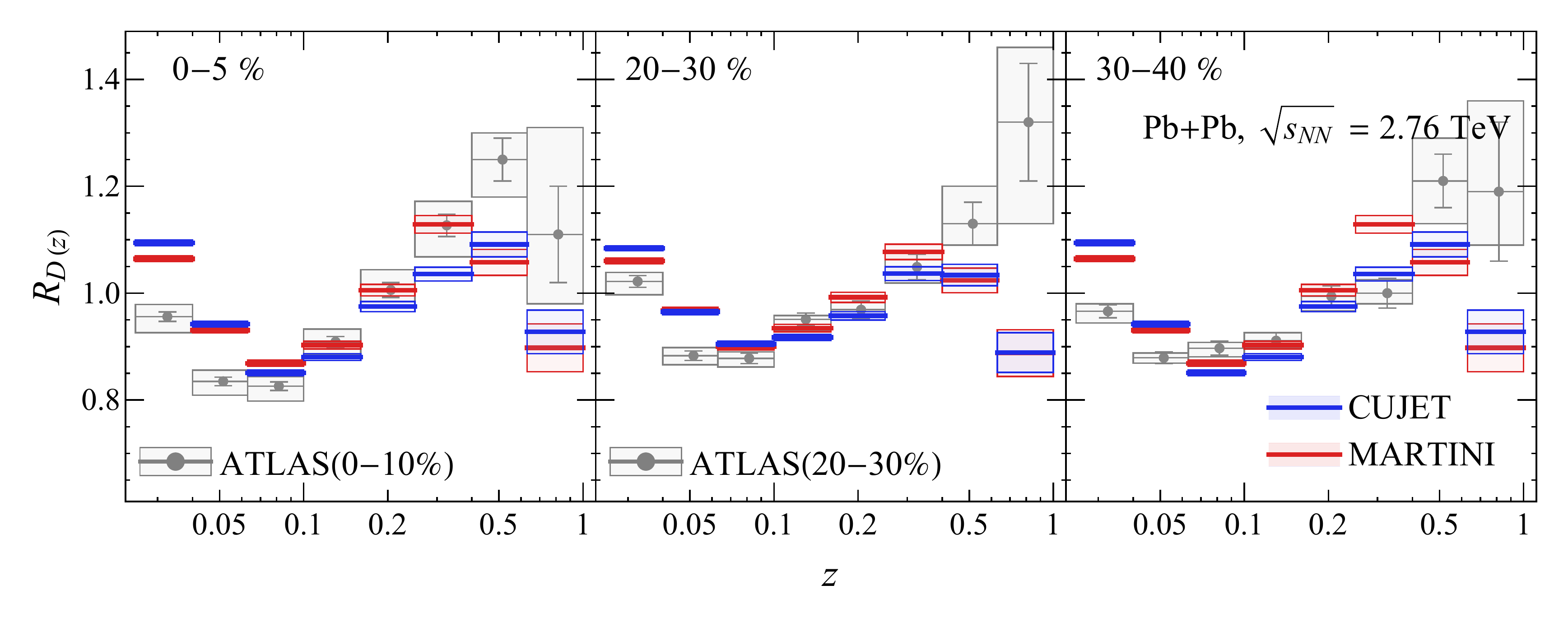} 
    \includegraphics[width=0.9\textwidth]{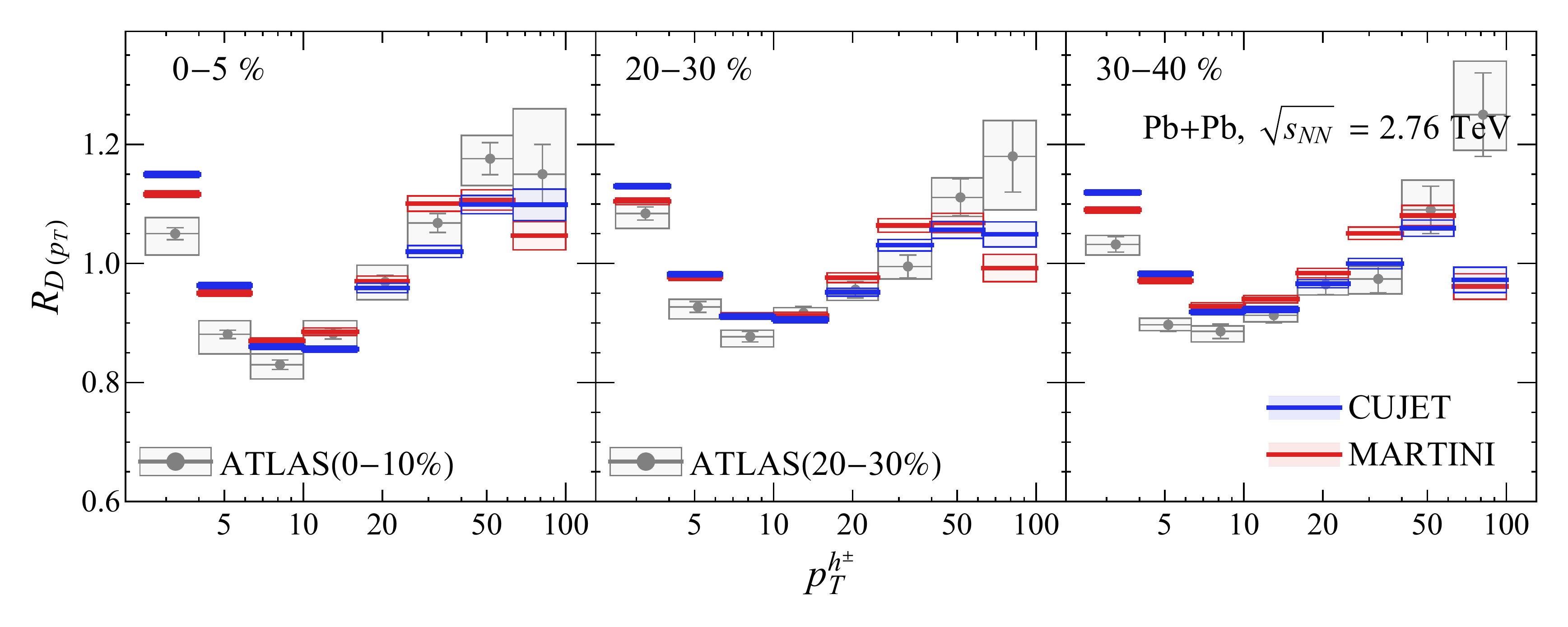}
    \caption{Fragmentation function ratio $R_{D}$ versus momentum fraction $z$ (top) and transverse momentum $p_T$ (bottom) for anti-$k_T$ $R=0.4$ jets with momentum $100<p_T^\mathrm{jet}<398$~GeV and pseudorapidity $|\eta_\mathrm{jet}|<2.1$. From left to right are simulation results are respectively for $0-5\%$, $20-30\%$, and $30-40\%$ most central $\mathrm{Pb}+\mathrm{Pb}$ collisions at beam energy $\sqrt{s_{NN}}=2.76$~TeV. ATLAS results~\cite{ATLAS:2017nre} are also presented for comparison.
    \label{fig.AA.fragmentationfunction}}
\end{figure*}

We begin with a brief discussion of our workflow and the models used in this work. In particular, we focus on the jet-quenching phenomena in $\mathrm{Pb}+\mathrm{Pb}$ at beam energy $\sqrt{s_{NN}} = 2.76$~TeV and consider three centrality classes, $0$-$5\%$, $20$-$30\%$, and $30$-$40\%$, with the hydro background provided by the \jetscape\, Collaboration~\cite{JETSCAPE:2021ehl}. 
The simulations used an event-averaged initial state generated via \trento\, followed by a \vishnu~(2+1)-dimensional boost 
invariant viscous relativistic hydrodynamic simulation 
with temperature dependent shear and bulk viscosities. The 
parameters of the initial state and hydro were used from a Bayesian analysis in Ref$.$~\cite{Bernhard:2019bmu}.

\begin{table}
\centering
\begin{tabular}{l l l l}
\toprule 
    Model & Parameter & Value & Note \\ 
\midrule
\multirow{3}{*}{Both} 
    & $N_c$ & $3$ & number of colors\\
    & $\Lambda_\mathrm{QCD}$ & $0.2~\text{GeV}$ & Eqs.~\protect{\eqref{eq.martini.running_coupling}, \eqref{eq.cujet.running_coupling}}\\
    & $p_{\mathrm{cut}}$ & $2.0~\text{GeV}$ & cut for energy loss \\
\midrule
\multirow{4}{*}{\martini}
    & $N_f$ & $3$ & number of flavors\\
    & $\alpha_{s,0}$ & $0.3$ & Eq.~\protect{\eqref{eq.running_scales_martini}}\\
    & $\kappa_r$ & $1.5$ & Eq.~\protect{\eqref{eq.running_scales_martini}}\\
    & $\kappa_e$ & $4.5$ & Eq.~\protect{\eqref{eq.running_scales_martini}}\\
\midrule
\multirow{2}{*}{\cujet}
    & $N_f$ & $2.5$ & number of flavors\\
    & $\alpha_\mathrm{max}$ & $0.3$ & Eq.~\protect{\eqref{eq.cujet.running_coupling}}\\
\bottomrule
\end{tabular}
\caption{A summary table of \martini\, and \cujet\, parameters used in this calculation. See also Table.~I of~\protect{\cite{JETSCAPE:2019udz}} for other parameters in the \jetscape\, framework.\label{table.model.parameters.used}}
\end{table}
   
The hard sector events are generated by first using \pythia\, to generate the hard scattering event 
with initial state radiation and multiparton interactions but no final state showers. The highly 
virtual partons coming out of the hard scattering encounter an expanding hydrodynamic medium, and as such we employ \matter\, to simulate their in-medium energy loss. \matter\, then handles the energy loss of any energetic parton which either has left the medium (with remaining virtuality) or has virtuality $Q>Q_{0}$ with $Q_0$ set to 
$2$ GeV~\cite{JETSCAPE:2021ehl}. Jet partons with $Q <Q_{0}$ are taken to be on the mass shell and passed to the low virtuality 
energy loss module, \cujet\, or \martini\, to be further evolved in the medium. We also use a momentum cut ($p_{\mathrm{cut}} = 2$ GeV) in both \martini\, and \cujet\, modules. Partons with momentum below this scale are not permitted to interact with the medium. This is needed as the assumption behind both models is that the incoming parton jet is much more energetic and therefore distinguishable from the medium particles around it. Once the evolution is completed, 
the event is hadronized using the \texttt{"colorless"} hadronization module of \jetscape. Finally, jet 
clustering is performed using the anti-$k_T$ jet finding algorithm~\cite{Cacciari:2008gp} of \fastjet3~\cite{Cacciari:2011ma,Cacciari:2005hq} and the results are binned according to the appropriate experimental cuts. We summarize the parameters used in Table~\ref{table.model.parameters.used}.

Given that no medium is present in a \pp\, collisions, there would be no low-virtuality energy loss via \cujet\, or \martini. Overall 
we adopt the tuned parameter set of Ref$.$~\cite{JETSCAPE:2019udz}. More details on our \pp\, calculation and 
the associated results are provided in Appendix.~\ref{sec.pp_baseline}. 

As mentioned previously, the \cujet\, model has one free parameter, $\alpha_\mathrm{max}$, which is 
the maximum cutoff of running coupling $\alpha_s(Q^2)$. In this work, we tune the coupling 
parameters separately for each model in order to match the experimental results for charged 
hadron nuclear modification factor for $0$-$5\%$  $\mathrm{Pb}+\mathrm{Pb}$ collisions with beam 
energy $\sqrt{s_{NN}}=2.76~\mathrm{TeV}$~\cite{ALICE:2012aqc,ATLAS:2015qmb,CMS:2012aa}. The 
nuclear modification factor is defined as
\begin{align}
    R_{AA}(p_T) \equiv \frac{\mathrm{d}\sigma_{AA}/\mathrm{d}p_T}{N_\text{bin}\,\mathrm{d}\sigma_{pp}/\mathrm{d}p_T}\,,
    \label{eq.raa}
\end{align}
where $p_T$ is the transverse momentum of the energetic hadron, and $N_\text{bin}$ the number of binary 
collision of the nucleus-nucleus collisions.

We obtained $\alpha_\mathrm{max}=0.68$ for \cujet. The free parameters in \martini\, for the running of the strong 
coupling, i.e., $\alpha_{s,0}=0.3$, $\kappa_{e}=4.5$, and $\kappa_{r}=1.5$, are obtained by fitting charged hadron 
and jet nuclear modification factors (\raa)~\cite{Park:2021yck}.

Figure.\ref{fig.AA.chRAA} shows the comparison of the two models using the resulting charged hadron nuclear 
modification factor~($R_{AA}^{h^\pm}$). With their parameters tuned separately, we find that these models 
result in the same charged hadron \raa\, for $p_T^{h^\pm} \gtrsim 10~\text{GeV}$ across different centrality 
bins, and both of them agree well with the experimental data.

We can now move on to the nuclear modification factor for inclusive jets, which is defined in the same way as Eq.~\eqref{eq.raa} but with $p_T$ representing the transverse momentum of jet. In our simulations, jets are 
reconstructed using the same criteria used by the experiments~\cite{ATLAS:2014ipv,CMS:2016uxf}, i.e. using anti-$k_T$ 
algorithm with cone size $R=0.4$. Both the charged and neutral particles are included in the reconstruction of 
jets with no cut placed on their transverse momenta. As observed in Fig. \ref{fig.AA.jetRAA}, we find $R_{AA}^\mathrm{jet}$ predicted by the two models to also agree with each other as well as with the experimental data. It is worth noting that we observe good agreement in the \cujet-to-\martini\, and model-to-data comparison for jet cone sizes $R=0.2$, $0.3$, and $0.4$, at various centrality ranges. Results can be found in Fig.~\ref{fig.AA.jetRAA_full} of Appendix~\ref{appendix_C}.

\begin{figure*}[!hbtp]
        \centering
        \includegraphics[width=0.9\textwidth]{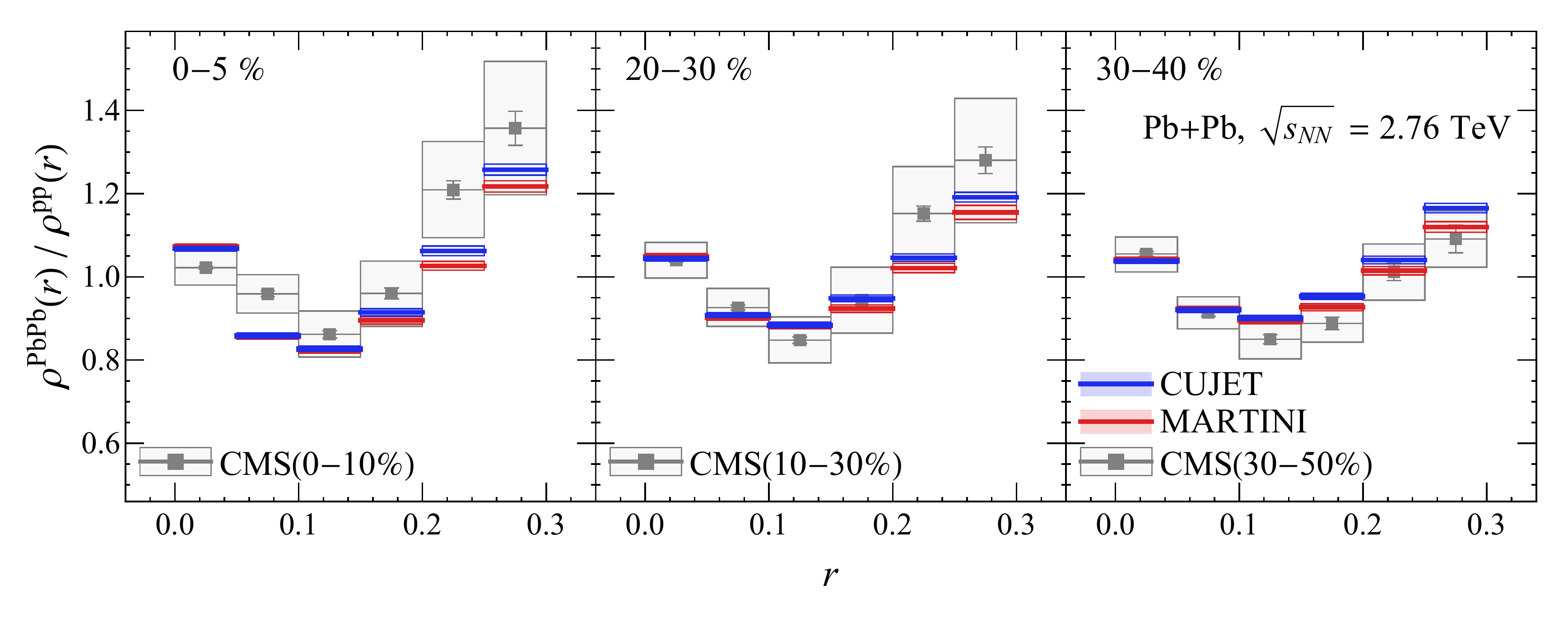}
        \caption{Jet shape ratio as a function of radial distance $r$ for jets within the kinematic region $p_T^\mathrm{jet}>100$~GeV and $0.3<|\eta_\mathrm{jet}|<2.0$. Jets are constructed using anti-$k_T$ 
        algorithm with radius $R=0.3$ with cut $p_T^\mathrm{trk}>1$~GeV. From left to right are simulation 
        results are respectively for $0-5\%$, $20-30\%$, and $30-40\%$ most central \PbPb\,
        collisions at beam energy $\sqrt{s_{NN}}=2.76$~TeV. CMS results~\cite{CMS:2013lhm} for $0-10\%$, 
        $10-30\%$, and $30-50\%$ centrality classes are also presented for comparison.
        \label{fig.AA.jetshape}}
        \includegraphics[width=0.9\textwidth]{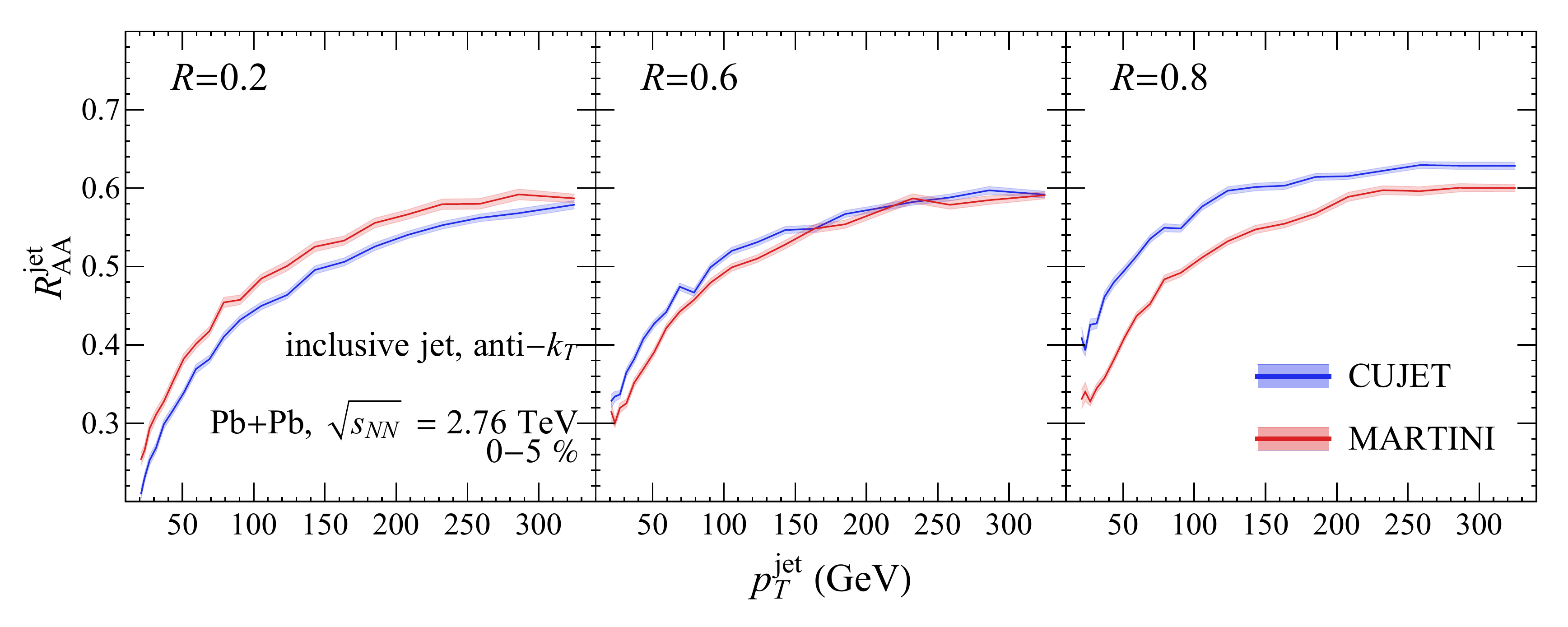}
        \caption{The same as Fig.~\protect{\ref{fig.AA.jetRAA}}~(left) but for jet cone size $R=0.2$, $0.6$ and $0.8$.
        \label{fig.AA.jetRAA_size}}
\end{figure*}
With the overall opacity, i.e., $\raa$ for charged hadron and jets, well described, we move on to study the medium modification of jet substructure characterized by the \PbPb\, to \pp\, ratio of the fragmentation function(FF) of charged hadrons in jets as well as the jet shape. The FF is used to describe the distribution of momentum fraction along jet axis for charged hadrons within a jet and is 
defined as
    \begin{align}\begin{split}
    	D(z)_{z \in [z_\mathrm{min},z_\mathrm{max})} 
          \equiv\;& 
        \frac{\sum_\mathrm{jets} \sum_{z \in [z_\mathrm{min},z_\mathrm{max})}1}{N_\mathrm{jet} \; (z_\mathrm{max} - z_\mathrm{min})}
    	 \,,
    \end{split}\end{align}
where $N_\mathrm{jet} \equiv  \sum_\mathrm{jets} 1$ is the total number of jets within the 
selected kinematic region, and $z$ is the charged hadron momentum fraction along the direction 
of the jet momentum:
    \begin{align}
        z \equiv \frac{\mathbf{p}_\mathrm{jet} \cdot \mathbf{p}_\mathrm{trk}}{\mathbf{p}_\mathrm{jet} \cdot \mathbf{p}_\mathrm{jet}} \,.
\end{align}
Similarly, one can define the FF with respect to the hadron transverse-momentum,
    \begin{align}\begin{split}
    	D(p_T)_{p_T \in [p_{T}^{\mathrm{min}}, p_{T}^{\mathrm{max}})} 
    \equiv\;& 
	    \frac{\sum_\mathrm{jets} \sum_{p_{T,\mathrm{trk}} \in [p_{T}^{\mathrm{min}},p_{T}^{\mathrm{max}})}1}{N_\mathrm{jet} \; (p_{T,\mathrm{max}} - p_{T,\mathrm{min}})}\,.
    \end{split}\end{align}

These two observables contain the same information when evaluated for a fixed momentum jet. They are different, however, in how they weight and bin the charged hadrons when averaging over jets with different energies.
By measuring the ratio of FF in $\mathrm{A}+\mathrm{A}$ collisions to that in \pp\, collisions,
    \begin{eqnarray}
	    R_{D(z)} &\equiv& D^\mathrm{AA}(z) / D^\mathrm{pp}(z) \,,\\
	    R_{D(p_T)} &\equiv& D^\mathrm{AA}(p_T) / D^\mathrm{pp}(p_T) \,,
    \end{eqnarray}
one can quantify and study the effect of in-medium fragmentation from each jet energy loss model. In the above equations, $\mathbf{p}_\mathrm{jet}$ and $p_{T,\mathrm{jet}}$ stand for momentum of the full jet, including both charged and neutral particles, although the triggering tracks are for charged particles only. This is in alignment with experimental measurements.

We show the \cujet\, and \martini\, results of fragmentation function ratio in Fig.~\ref{fig.AA.fragmentationfunction}, together with the ATLAS results~\cite{ATLAS:2017nre}. Although 
the difference between simulation results are within a factor of 2 of the statistical uncertainty, one can 
observe the systematic trend that $R_D^{\cujet}(z>0.7) > R_D^{\martini}(z>0.7)$ whereas 
$R_D^{\cujet}(z\approx0.3) < R_D^{\martini}(z\approx0.3)$. These trends agree with those of the splitting rates shown in the 
lower panels of Fig.~\ref{fig.rate}: for jets with initial momentum $p_\mathrm{ini}=100$~GeV, the \cujet\,
mechanism expects fewer splittings, especially for the range such that $p_g/p_\mathrm{ini}\gtrsim0.3$. The 
difference in splitting rate leads to the difference in the in-medium parton distribution, which is 
finally measured by the in-medium fragmentation function.

The jet shape observable is defined as
    \begin{equation}
        \rho(r) \equiv 
	        \frac{N_\mathrm{norm}}{N_\mathrm{jet}} \frac{\sum_\mathrm{jets} \sum_{r \in [r_\mathrm{min},r_\mathrm{max})} {p_{T}^{\mathrm{trk}}}/{p_{T}^{\mathrm{jet}}}}{r_\mathrm{max} - r_\mathrm{min}}\,,
    \end{equation}
to measure the charge hadron energy distribution along the angular distance perpendicular to the jet axis,
    \begin{equation}
        r \equiv \sqrt{(\phi_\mathrm{trk} - \phi_\mathrm{jet})^2 + (y_\mathrm{trk}-y_\mathrm{jet})^2} \,.
    \end{equation}

In Fig.~\ref{fig.AA.jetshape} we compare the jet shape ratio from both models, and CMS results~\cite{CMS:2013lhm} 
are also presented for comparison. While both \cujet\, and \martini\, are in broad agreement with the
experimental result given the large uncertainties, we can again observe 
systematic differences between the two. The \martini\, jet shape ratio is higher in the $r<0.05$ bin whereas the 
\cujet\, ratios are higher for bins with $r>0.05$ as we move away from the jet axis. This forms a complete story, when combined with the 
cone-size dependent jet \raa, which is shown in Fig.~\ref{fig.AA.jetRAA_size} and exhibits a flip in relative position between \martini\, and \cujet\, as we go from small to large jet cone radii.
In the radiation processes, the opening angle of 
the daughter particles is always small, hence both models assume that the outgoing gluon is emitted 
collinearly with the incoming parton. The medium modification of the radial shape of jets 
is dominated by elastic scatterings between the emitted gluons and the medium particles. As we saw 
in Sec$.$~\ref{sec.comparison.static}, \martini\, radiates many more soft gluons than \cujet. Soft 
gluons have a higher chance of being deflected via elastic scatterings with the medium.
Thus we observe more particles within the opening angle\footnote{note that
$0.8~\text{rad}=45.8^{\circ}$} $0.1\lesssim r \lesssim 0.8$ in \cujet, which is reflected by the jet shape and the cone-size dependent
jet-\raa. In contrast, the fact that $\raa^{\text{jet},R=0.2}[\cujet] < \raa^{\text{jet},R=0.2}[\martini]$ may be due to the stronger elastic scattering --- given the fitted value for the strong coupling --- and more deflection of the hard parton.

\section{Conclusion and Outlook}
\label{sec.conclusion.outlook}

Jet quenching phenomena in heavy-ion collisions provide a great opportunity for tomographic studies of the QGP medium that is created. Many models have been proposed, with various assumptions about the mechanisms of jet-medium interactions, to study and simulate jet quenching, and they have been quite successful at reproducing experimental observations. With the field now moving towards precision studies there is a need for more detailed, direct, and fair comparison and analysis of the quenching models. In this work in particular, we focused on a comparative study of \cujet\, and \martini\, formalisms for low virtuality jet energy loss. 
The former keeps diagrams up to the first order in opacity expansion and accounts for the finite medium size, whereas the latter is evaluated to all orders in opacity expansion and assumes infinite medium size. These two models are fundamentally different in the energy, temperature, and path length dependence of the radiative energy loss rates, and predict different momentum distributions for the radiated gluons. Indeed the result of our calculation in a QGP brick (Fig$.$~\ref{fig.radel}) clearly demonstrates this fact. Furthermore, previous ``standalone'' realistic simulations of jet quenching using \cujet\, and \martini\, were found to provide a good description of the high momentum charged hadron and jet \raa. Thus a fair comparative study within a realistic simulation, with as many parameters held fixed as possible was in order.

The \jetscape\ framework is a simulation package that provides state of the art components/models in simulating high-energy observables in heavy-ion collisions, including unquenched parton distribution, hydrodynamic background, hadronization, and jet clustering. Then, one can take different jet energy loss models and compute the high-energy observables while maintaining control over all other aspect of the simulation. In this work, we integrated the \cujet\ energy loss model into the \jetscape\ framework by first recasting it into a stochastic version for the first time, and then using it to calculate jet-quenching observables. The results of this calculation were then compared to those generated separately using \martini. This modularity allows for the use of jet physics in heavy ion collisions not just to study the thermal medium but also as a way to compare various models to each other. Thus leveraging the power of \jetscape\, enables us to learn more about the jet quenching models.

We focus on $\sqrt{s_{NN}}=2.76$~TeV \PbPb\, collisions and use the charged hadron \raa\, at $0$-$5\%$ centrality to tune the parameters of the two models. Not much difference can be observed between the two when comparing charged hadron or jet \raa. The differences between the formalisms begin to manifest themselves when considering the fragmentation function ratios~(Fig$.$~\ref{fig.AA.fragmentationfunction} top) where the models are very close to each other along the jet momentum axis but populate the hadrons with lower energy fractions~($z$) within the jet differently. Similar behavior is observed in FF ratio as a function of charged hadron $p_T$ where \cujet\, and \martini\, predict different distributions of charged hadrons, particularly in the most central events, but ones that are still compatible with data within experimental uncertainties. While fragmentation function ratios are sensitive to the details of the radiative energy loss channel, the jet shape ratio results are more sensitive to the combination of radiative and collisional energy loss. This is evident in Fig$.$~\ref{fig.AA.jetshape} where along the jet axis the two models are nearly identical, and going away from the jet axis, across the three centrality classes considered here, \martini\, results lie slightly below those from  \cujet. This is the consequence of \martini's propensity for soft gluon radiation and their subsequent deflection due to elastic interactions with the medium. The flip in the ordering of the jet \raa\, observed in Fig$.$~\ref{fig.AA.jetRAA_size} further emphasizes this effect. Precise experimental measurements of jet \raa\, as a function of jet cone radius would then present an interesting opportunity and test of these models.

Another interesting and important factor to note is the large $\alpha_{s,\mathrm{max}}$ that is preferred by \cujet\, $\chi^2$ fit to charged hadron \raa. In particular, this large value is acquired when the formation time in Eq$.$~\eqref{eq.splitting_rate_cujet} is set to be the time since the last splitting rather than the current proper time. As such the suppression from the LPM phase is proportional to $\tau^2$, a small number which when fitting for charged hadron \raa\, needs to be compensated by a large $\alpha_{s,\mathrm{max}}$. For more details, see Appendix.~\ref{appendix_A}.

We end by noting that, similarly to the gluon radiation process ($q \to q + g$), the \dglv\ and \amy\ formalisms predict fundamentally characteristically different splitting rates for the bremsstrahlung photon production ($q \to q + \gamma$). While one can hardly measure the distribution of the radiated gluon, the bremsstrahlung photons can be directly observed in the final state. Therefore, these two models predict different direct photon spectra in heavy ion collisions, which may be an independent discriminator of the energy loss mechanisms. This will be investigated in our followup paper.

\vspace*{0.3cm}
\acknowledgments{The authors thank Dr. Chanwook Park for his help during the early stages of this work. We are grateful to L.~Du, P.~Jacob, A.~Kumar, A.~Majumder, C.~Shen, G.~Vujanovic, and B.~Zhang for helpful discussions. 
This work was funded in part by the Natural Sciences and Engineering Research Council of Canada, and in part by the U.S. Department of Energy, Office of Science, Office of Nuclear Physics, under Grants No.
DE-FG88ER41450 and No. DE-SC0012704. S.S. is grateful for support from Le Fonds de Recherche du Qu\'ebec - Nature et technologies (FRQ-NT), via a Bourse d'excellence pour \'etudiants  \'etrangers (PBEEE). Computations were made on the B\'eluga, Graham and Narval computers managed by Calcul Qu\'ebec and by the Digital Research Alliance of Canada. }
\bibliographystyle{apsrev4-1.bst}
\bibliography{Ref.bib}

\clearpage
\begin{appendix}
\section{Implementation of the CUJET energy loss model in the JETSCAPE framework}
\label{appendix_A}
In this work, we employ the \jetscape\, (version 2.0) simulation framework to sample energy loss according to both \amy-\martini\, and \dglv-\cujet\, energy loss models. In a realistic simulation, the \jetscape\, framework generates the initial parton distribution using \pythia\, with final-state radiation turned off; then the splitting of high-virtuality partons is simulated by \matter, whereas one can select one of the built-in models or implement their own model to simulate the radiative and collisional in-medium energy losses; finally the hadronization of partons is realized by the colorless hadronization model. The \jetscape\, framework also contains \fastjet, to construct jets out of final state hadrons.

In the official \jetscape\, framework, there are three built-in energy loss models: \martini~\cite{Schenke:2009gb}, \textsc{ads-cft}~\cite{Casalderrey-Solana:2014bpa}, and \textsc{lbt}~\cite{Luo:2018pto}. These models sample the in-medium elastic collision and inelastic radiations for hard partons in a Monte Carlo manner. While being similar in the elastic collision channels, they take different Ans\"{a}tze to simulate Landau--Pomeranchuk--Migdal effect for inelastic radiations. We integrated the \cujet\, model of parton energy loss into the \jetscape\, version 2.0 workflow as a low-virtuality energy loss module. The original standalone-\cujet\, simulation is a deterministic model and to incorporate it into a \jetscape\, workflow we need to cast it to a Monte Carlo simulation. 

In the \cujet\, module, we employ the radiation kernel in Eq.~\eqref{eq.splitting_rate_cujet} which leads to the following radiation probability within a small time interval $\delta\tau$:
\begin{align}
P_\mathrm{rad}^{i}(p,T,\tau_\mathrm{form}) = \delta\tau \int_0^1 \mathrm{d}z \frac{\mathrm{d}\Gamma^{\dglv}_{i \to g i}}{\mathrm{d}z}(p,T,z,\tau_\mathrm{form})\,,
\label{eq.rad_prob_cujet}
\end{align}
where $p$ and $T$ are respectively the momentum of the hard parton and the fluid temperature.
If radiation happened, the momentum of an out going radiated gluon is sampled accordingly, under the assumption of collinear emission which is valid for bremsstrahlung of ultra-relativistic particles up to leading order in the coupling constant. 
Regarding the elastic scatterings with the medium, we take the total cross section in Eq.~\eqref{eq.cujet.elastic} to compute the total collision probability for a small time interval $\delta\tau$:
\begin{align}
P_\mathrm{ela}^{i}(p,T) = \Gamma^{\cujet}_\mathrm{ela}(p,T)\, \delta\tau \,,
\end{align}
Then, the differential cross section~\eqref{eq.elastic_diff} is used to sample the microscopic configuration, i.e. momenta of in-coming and out going particles.

\begin{figure}[!hbtp]\centering
\includegraphics[width=0.45\textwidth]{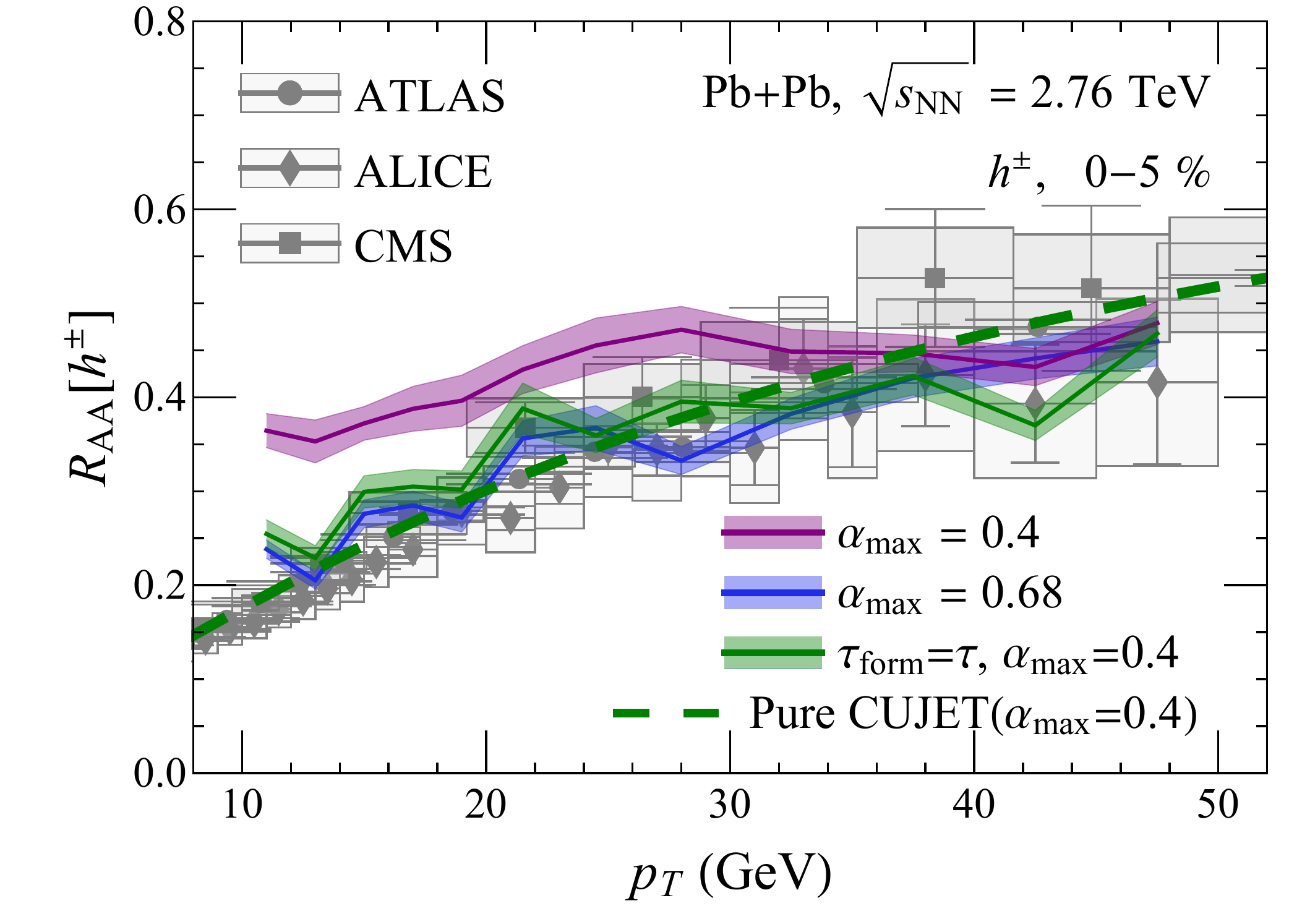}
\caption{Nuclear modification factor $R_{AA}$ versus transverse momentum $p_T$ for charged hadrons. Simulation results are for $0$-$5\%$ $\mathrm{Pb}+\mathrm{Pb}$ collisions at beam energy $\sqrt{s_{NN}}=2.76$~TeV. See text for explanation of different simulation results. Experimental results from the ALICE~\cite{ALICE:2012aqc}, ATLAS~\cite{ATLAS:2015qmb} and CMS~\cite{CMS:2012aa} Collaborations are also presented for comparison.
\label{fig.AA.cujet_test}}
\end{figure}
In Eq.~\eqref{eq.rad_prob_cujet}, the radiation probability explicitly depends to time ($\tau_\mathrm{form}$) accounting for the LPM effect. In our implementation of the \dglv\, energy-loss model, $\tau_\mathrm{form}=\tau-\tau_\mathrm{last\,split}$  is the difference between the current proper time and the latest time that such a partons radiates a gluon. On the other hand, the standalone-\cujet\, framework~\cite{Buzzatti:2011vt,Xu:2014ica,Xu:2014tda,Xu:2015bbz,Shi:2018izg,Shi:2018lsf} computes the jet-energy loss deterministically by taking the trajectory integrals and therefore we always let $\tau_\mathrm{form}$ be the current proper time. To investigate the influence of time definition, we fix the strong coupling with $\alpha_\mathrm{max} = 0.4$ and run simulations in the \jetscape\, framework using two definitions: $\tau_\mathrm{form}=\tau-\tau_\mathrm{last\,split}$ and $\tau_\mathrm{form}=\tau$. Results of charge hadron nuclear modification factor are respectively represented by purple and green bands in Fig.~\ref{fig.AA.cujet_test}. Corresponding experimental measurements~\cite{ALICE:2012aqc,ATLAS:2015qmb,CMS:2012aa} as well as original \cujet\, results~\cite{Shi:2018izg}, with the same coupling constant, are also shown.
Using the former definition of $\tau_\mathrm{form}$, the second gluon radiation would be further suppressed and the high energy particles are less quenched. In order to reach the same amount of energy loss, one would need to increase the coupling. From a $\chi^2$ fit of the charged hadron nuclear modification factor in $0$-$5\%$ $2.76$~TeV $\mathrm{Pb}+\mathrm{Pb}$ collisions, we found $\alpha_\mathrm{max} = 0.68$. It is worth noting that the Monte Carlo simulation using the same coupling and the same definition of $\tau_\mathrm{form}$ as the deterministic standalone \cujet\, leads to the same $R_{AA}$ as the latter (see green band and dash curve), which indicates the consistency between these two approaches.

\section{$\mathrm{p}+\mathrm{p}$ results in the JETSCAPE framework}
\label{sec.pp_baseline}
	\begin{figure*}[!hbt]\centering
		\includegraphics[width=0.36\textwidth]{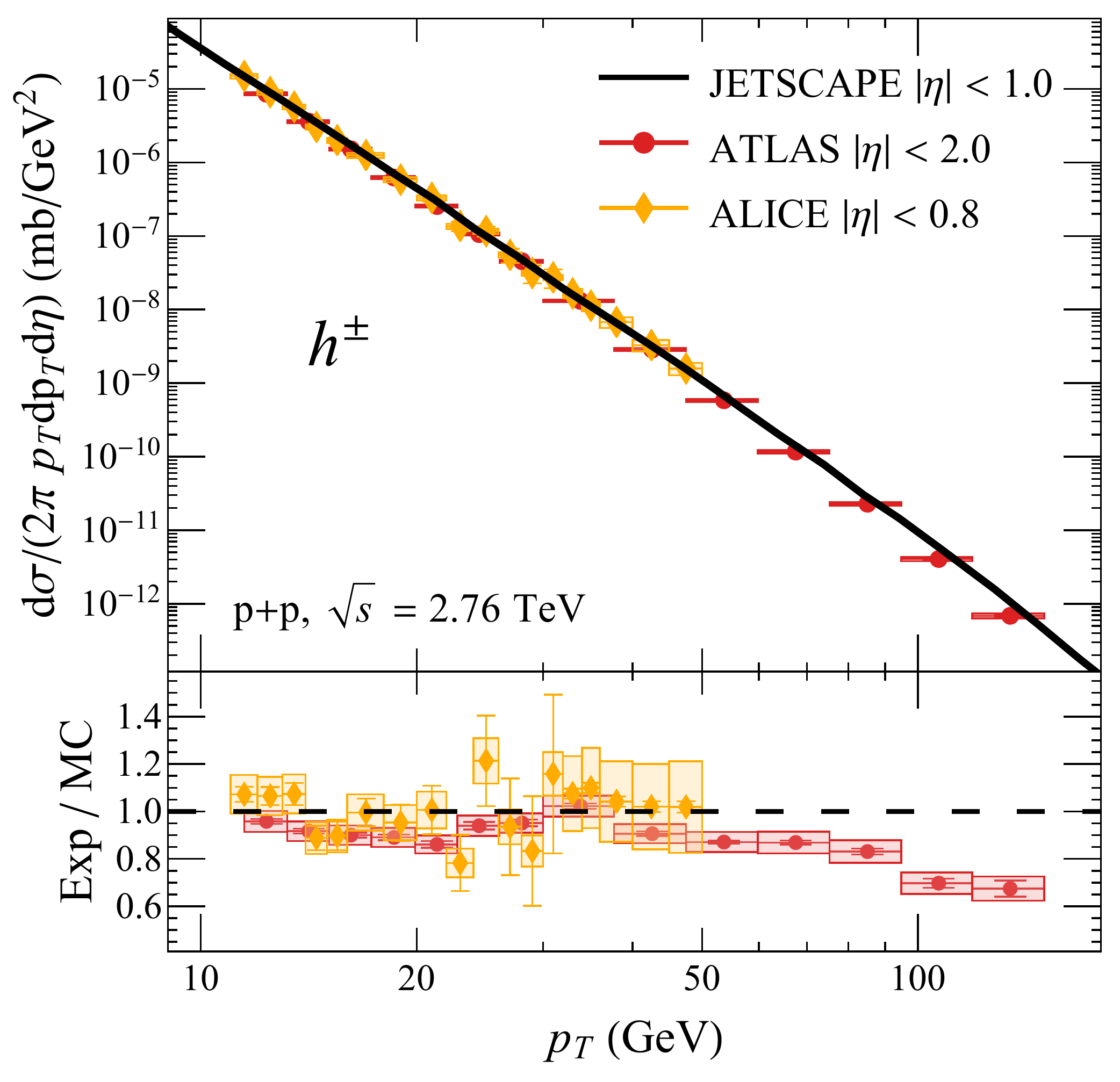} \qquad
		\includegraphics[width=0.36\textwidth]{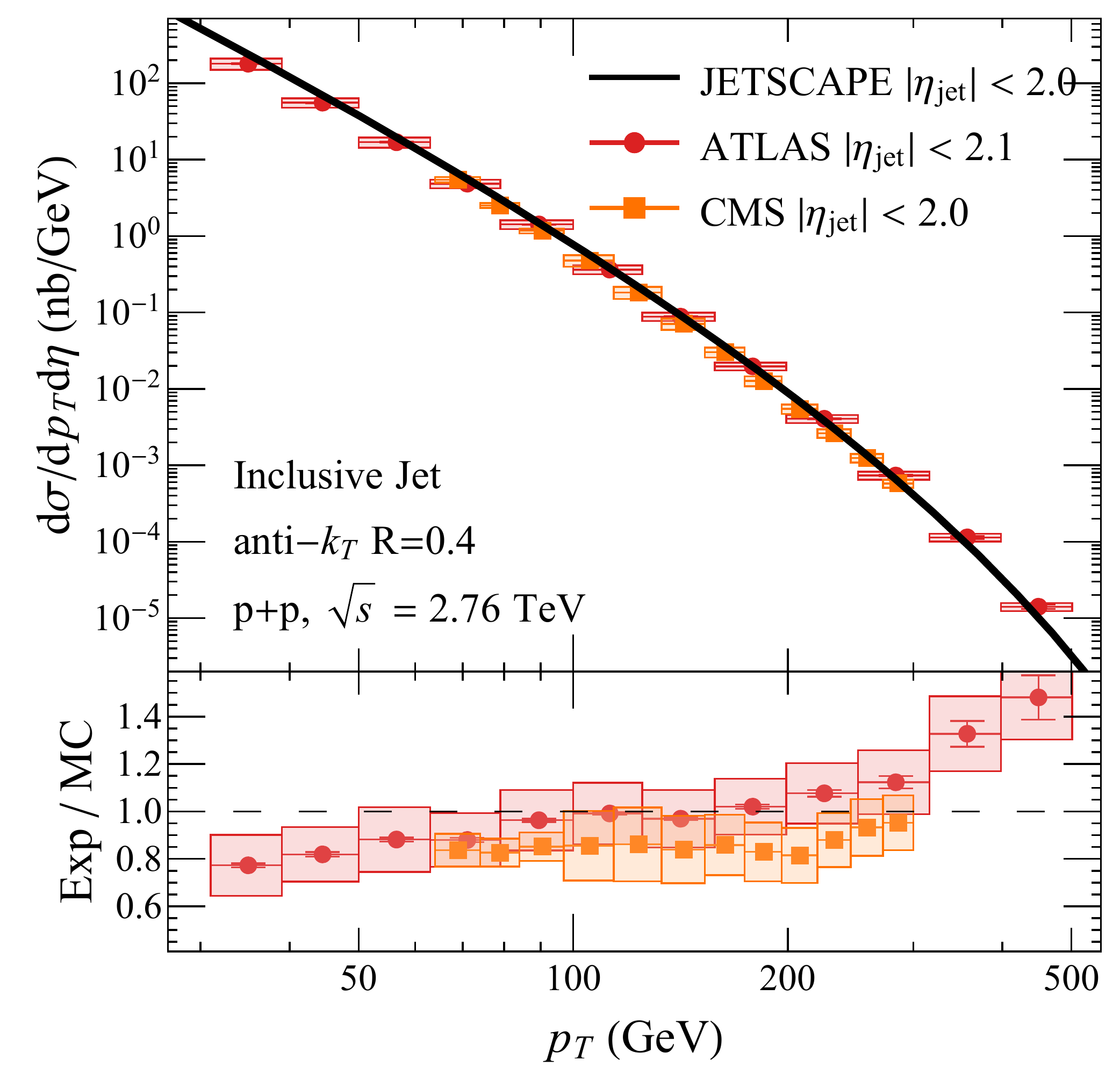}
		\caption{Charged hadron production cross section (left) and inclusive jet production cross section (right) for $\mathrm{p}+\mathrm{p}$ collisions at beam energy $\sqrt{s}=2.76$~TeV. Black curves are \jetscape\, results. Experimental data for charged hadron (inclusive jet) cross-section are from ATLAS~\cite{ATLAS:2015qmb} and ALICE~\cite{ALICE:2013txf} (ATLAS~\cite{ATLAS:2014ipv} and CMS~\cite{CMS:2016uxf}) Collaborations. For both simulation and experimental results, jets are constructed using the anti-$k_T$ algorithm with jet radius $R=0.4$.
		\label{fig.pp.crosssection}}
		\vspace{3mm}
		\includegraphics[width=0.32\textwidth]{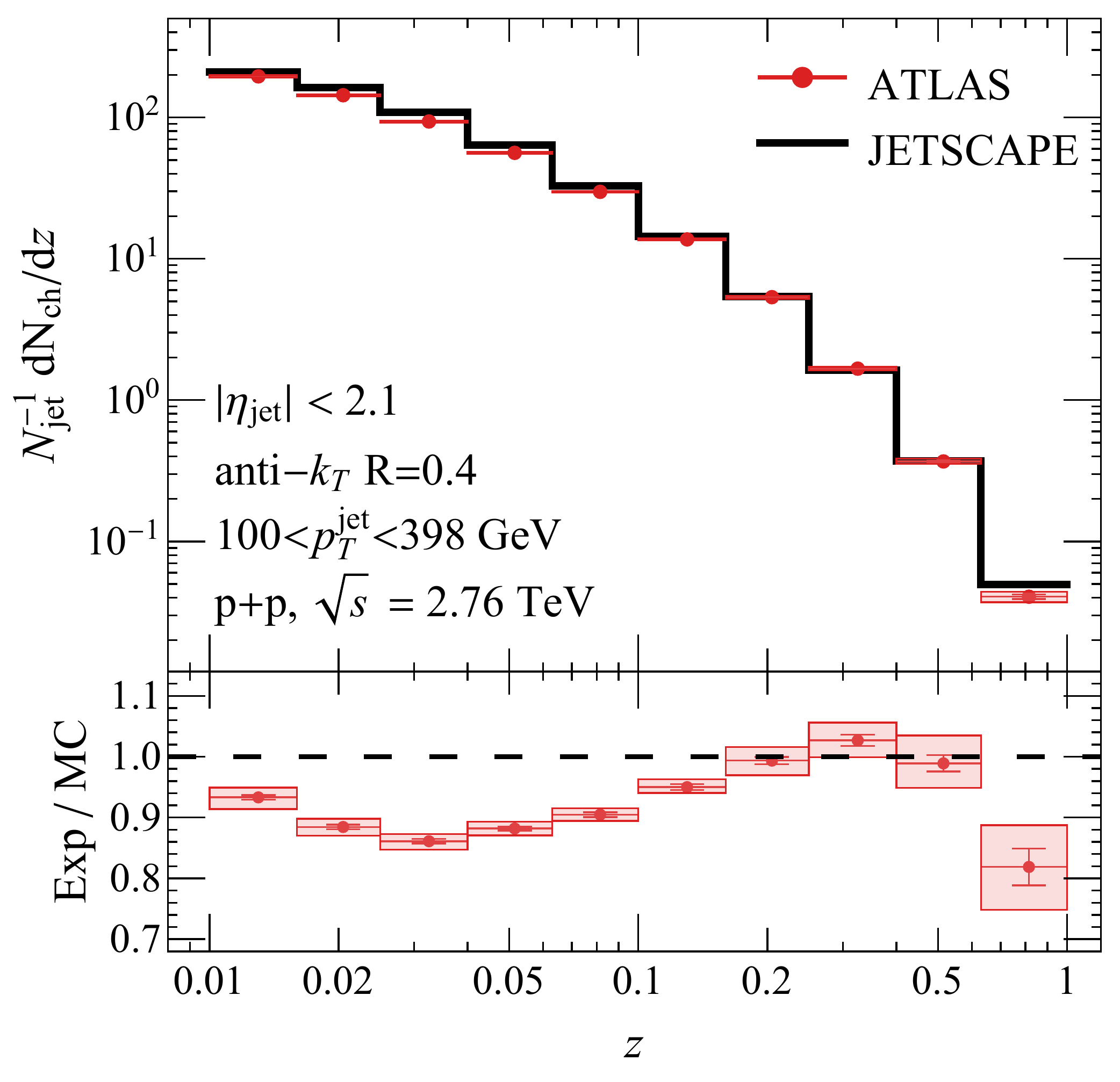} 
		\includegraphics[width=0.32\textwidth]{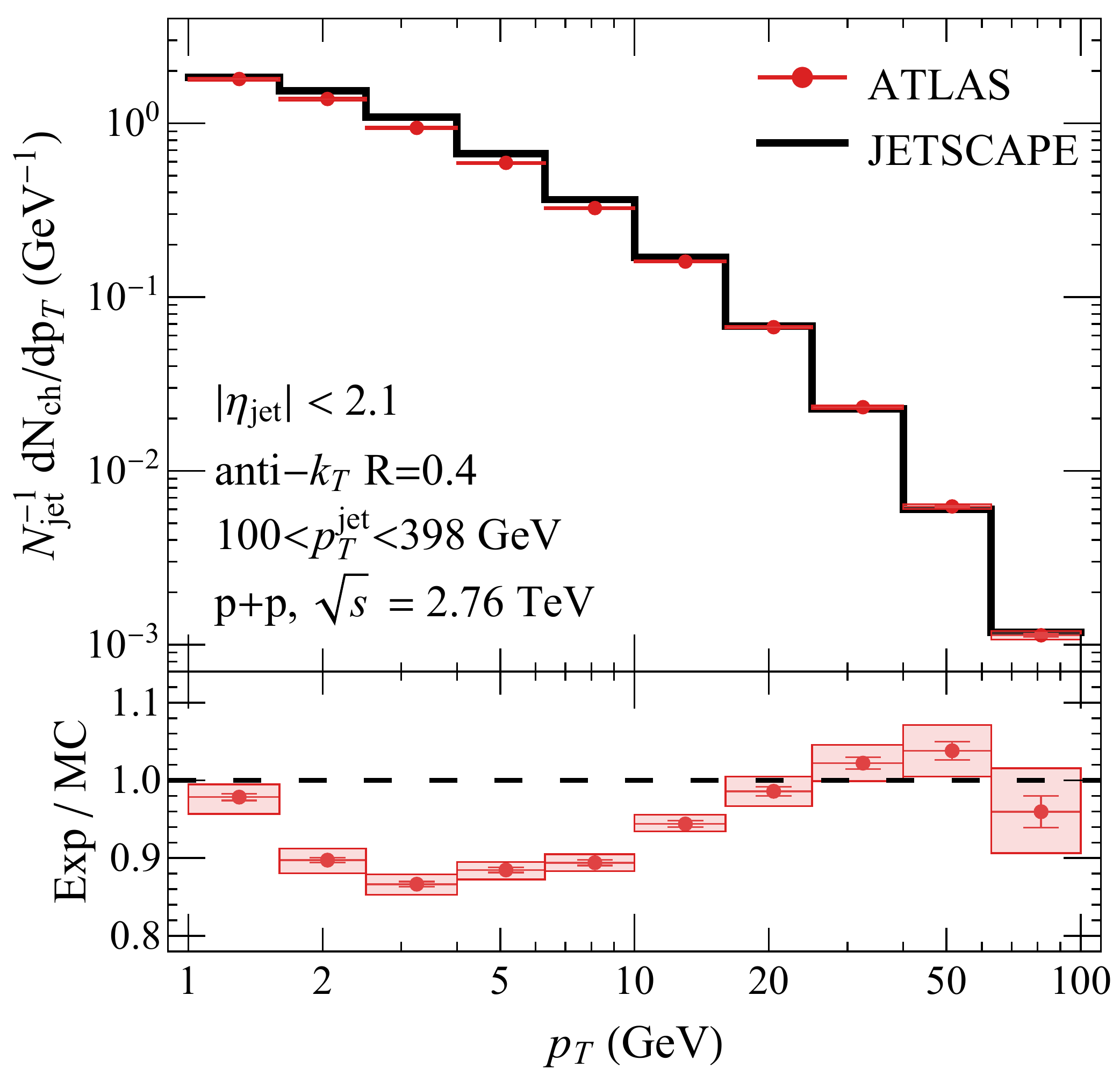}
		\includegraphics[width=0.32\textwidth]{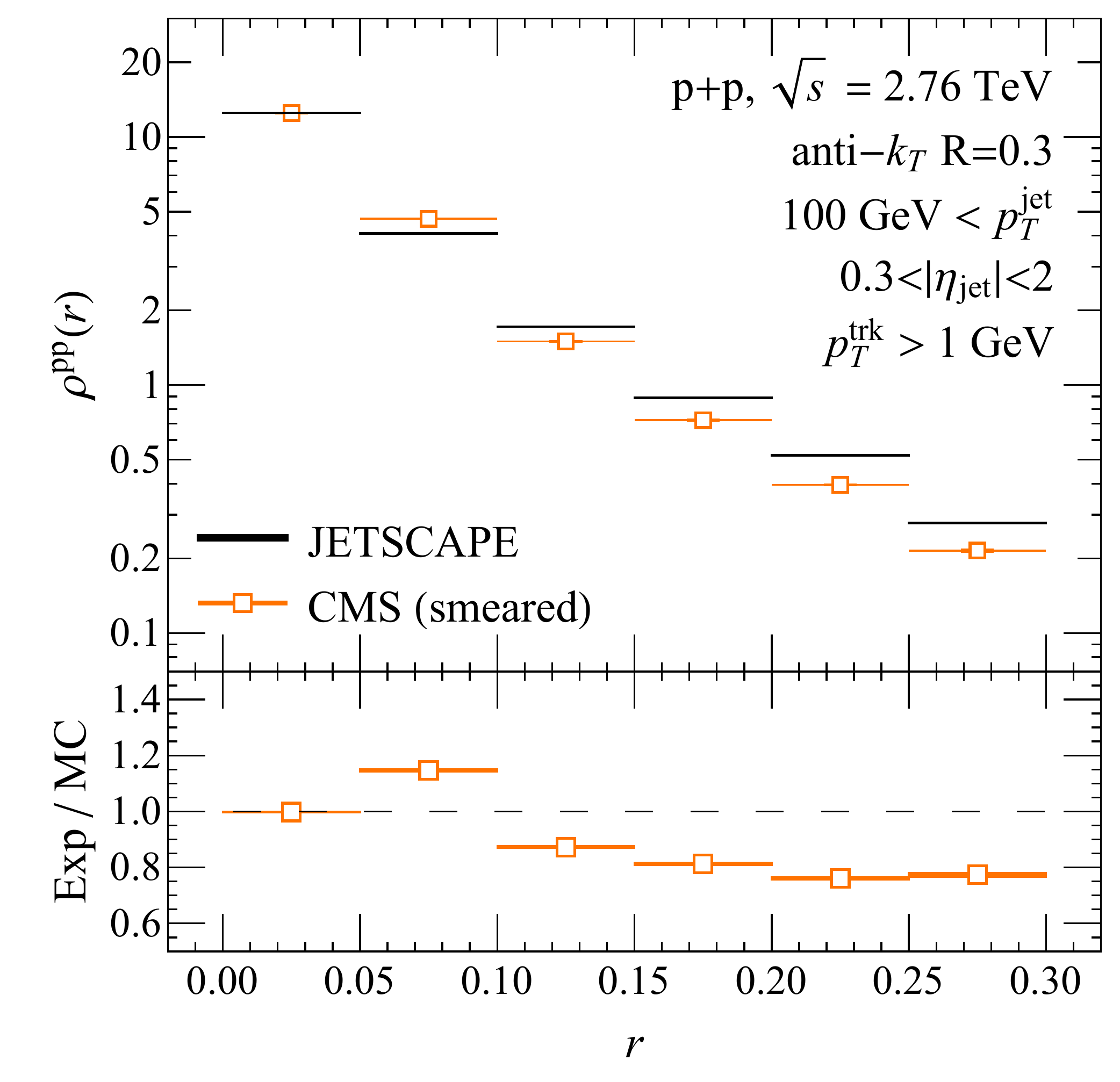}
		\caption{Jet substructure observables for $\mathrm{p}+\mathrm{p}$ collisions at beam energy $\sqrt{s}=2.76$~TeV. 
Left and middle panels are respectively for the jet fragmentation function versus momentum fraction $z$ and momentum $p_T$, and the right panel shows jet shape versus distance $r$.
Black curves are \jetscape\, results, while red circles (orange squares) are experimental data from the ATLAS~\cite{ATLAS:2017nre} (CMS~\cite{CMS:2013lhm}) Collaboration. 
Jet reconstruction techniques and criteria are the same as in the corresponding experimental analysis: for fragmentation functions, jets are constructed using the anti-$k_T$ algorithm with jet radius $R=0.4$ in the kinematic region such that $100<p_T^\mathrm{jet}<398$~GeV and $|\eta_\mathrm{jet}|<2.1$, whereas, for jet shape, jets are reconstructed using the anti-$k_T$ algorithm with jet radius $R=0.3$, with $p_T^\mathrm{jet}>100$~GeV and $0.3<|\eta_\mathrm{jet}|<2$.
		\label{fig.pp.substructure}}
	\end{figure*}

In order to provide reliable calculations of medium modification effects, one has to first ensure agreement with $\mathrm{p}+\mathrm{p}$ experimental data.
In this work, we adopt the default \jetscape\, setup which is tuned for $2.76$~TeV $\mathrm{p}+\mathrm{p}$ collisions~\cite{JETSCAPE:2019udz}.
Particularly, for fair comparison with $\mathrm{Pb}+\mathrm{Pb}$ simulation to reveal the medium modification, final-state-radiation in \pythia\, is turned off and the splitting of off-shell initial partons (virtuality $Q_0>1$~GeV) is described by \matter\, vacuum radiation, i.e. in-medium and recoiling effects are turned off.
In this section we show high $p_T$ observables provided by \jetscape\, simulation for $\mathrm{p}+\mathrm{p}$ collisions with beam energy $\sqrt{s}=2.76~\mathrm{TeV}$.

We compare simulation and experimental results for the production cross section of charged hadrons~\cite{ATLAS:2015qmb,ALICE:2013txf} (Fig.~\ref{fig.pp.crosssection} left) and inclusive jets~\cite{ATLAS:2014ipv,CMS:2016uxf} (Fig.~\ref{fig.pp.crosssection}, right), the fragmentation functions at different $z$ bins (Fig.~\ref{fig.pp.substructure}, left) and $p_T$ bins~\cite{ATLAS:2017nre} (Fig.~\ref{fig.pp.substructure} middle), and finally the jet shape~\cite{CMS:2013lhm} (Fig.~\ref{fig.pp.substructure}, right).
We employ the same technique and criteria as experimental analysis to reconstruct and select jets. Generally speaking, we employ the anti-$k_T$ algorithm with an appropriate choice of jet-cone radius $R$ to construct jets from final state stable particles, including charged and neutral particles with specific kinematic criteria to match the corresponding measurements.
In the jet cross section comparison (Fig.~\ref{fig.pp.crosssection} right), we take $R=0.4$ and select jet events with $|\eta_\mathrm{jet}|<2.0$.
For fragmentation function measurement (Fig.~\ref{fig.pp.substructure}, left and middle), we take $R=0.4$ and select jet events with $|\eta_\mathrm{jet}|<2.1$, $p_T^\mathrm{jet}\in[100,398]$~GeV.
For jet shape measurement (Fig.~\ref{fig.pp.substructure}, right), we take $R=0.3$, $p_T^\mathrm{trk}>1$~GeV and select jet events with $0.3<|\eta_\mathrm{jet}|<2.0$, $p_T^\mathrm{jet}>100$~GeV. Across all these observables, we see satisfactory performance for our $p$-$p$ calculations to be used as the baseline for the heavy-ion simulations.

\begin{widetext}
\section{Centrality and Cone-Size Dependence of Jet Nuclear Modification Factor}\label{appendix_C}

In this appendix we show the simulation results for nuclear modification factor of jets with various cone sizes and at various centrality classes. For the centrality ranges not provided by the official \jetscape\ package, we generate them using the same code and same parameter setting~\cite{Bernhard:2019bmu} as used by the \jetscape\ Collaboration. Result are shown in Fig.~\ref{fig.AA.jetRAA_full}, and we observe good agreement with the CMS data~\cite{CMS:2016uxf}.
\begin{figure}[!hp]
    \centering
    \includegraphics[width=\linewidth]{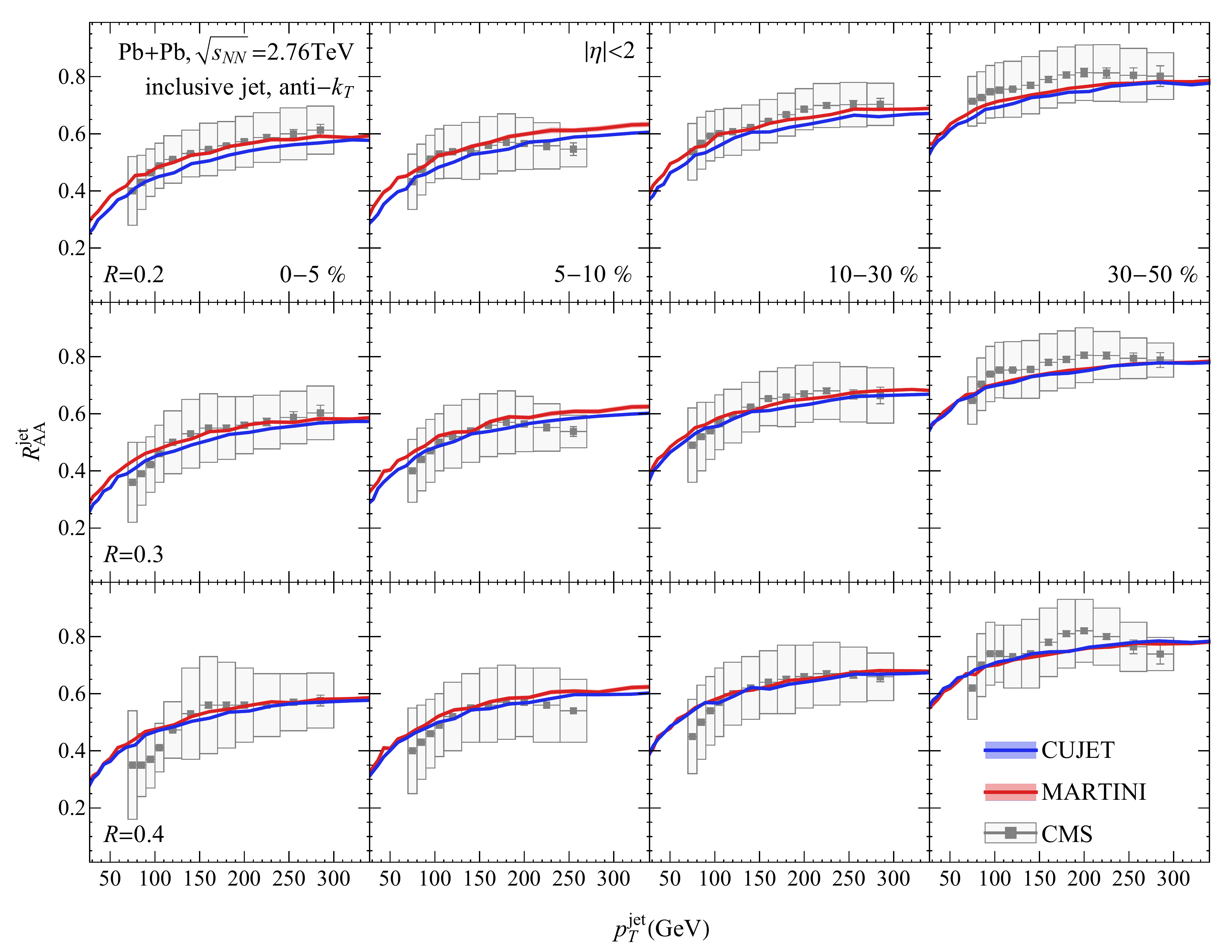}
    \caption{Same as Fig.~\protect{\ref{fig.AA.jetRAA}} but for $0$-$5\%$, $5$-$10\%$, $10$-$30\%$, and $30$-$50\%$ centrality classes and with cone sizes $R=0.2$, $0.3$, and $0.4$.
    \label{fig.AA.jetRAA_full}}
\end{figure}
\end{widetext}
\end{appendix}
\end{document}